\documentclass[12pt]{article}
\usepackage{epsf}
\usepackage{latexsym}
\usepackage{bm}
\DeclareMathAlphabet{\mib}{OML}{cmm} {b}{it}
\def\b{\beta}

\def\th{\theta}

\newcommand{\g}{\gamma}

\newcommand{\del}{\partial}

\newcommand{\half}{{1 \over 2}}
\newcommand{\st}{{\tilde t}}

\newcommand{\absv}[1]{\left|#1\right|}

\newcommand{\gtsim}{\mathrel{\hbox{\raise0.2ex
\hbox{$>$}\kern-0.75em\raise-0.9ex\hbox{$\sim$}}}}
\newcommand{\ltsim}{\mathrel{\hbox{\raise0.2ex
\hbox{$<$}\kern-0.75em\raise-0.9ex\hbox{$\sim$}}}}
\newcommand{\lw}[1]{\smash{\lower2.0ex\hbox{#1}}}

\newcommand{\Tr}{{\rm Tr}}

\newcommand{\kslash}{k\kern-0.5em\raise 0.14ex\hbox{/}}
\newcommand{\PRD}[3]{Phys. Rev. {\bf D{#1}} (19{#2}) {#3}}
\newcommand{\PRLet}[3]{Phys. Rev. Lett. {\bf {#1}} (19{#2}) {#3}}
\newcommand{\NPB}[3]{Nucl. Phys. {\bf B{#1}} (19{#2}) {#3}}
\newcommand{\PLB}[3]{Phys. Lett. {\bf B{#1}} (19{#2}) {#3}}
\newcommand{\PrTP}[3]{Prog. Theor. Phys. {\bf {#1}} (19{#2}) {#3}}
\makeatletter

\@addtoreset{equation}{section}
\makeatother
\begin{document}
\begin{titlepage}
\begin{flushright}
SAGA--HE--141\\
September 24,~1998
\end{flushright}
\vspace{50pt}
{\Large\bf
\begin{center}
Higgs Mass, $\mib{CP}$ Violation and Phase Transition\\
in the MSSM
\end{center}}
\vskip2.0cm
\begin{center}
{\bf Koichi~Funakubo\footnote{e-mail: funakubo@cc.saga-u.ac.jp}}
\end{center}
\vskip 1.0 cm
\begin{center}
{\it Department of Physics, Saga University,
Saga 8408502 Japan}
\end{center}
\vskip 1.5 cm
\centerline{\bf Abstract}
\vskip 0.2 cm
\baselineskip=15pt
The effective potential in the MSSM at the one-loop level is used to
evaluate masses of the neutral Higgs scalars and to study 
finite-temperature phase transition.
The $CP$ violation in the Higgs sector,
which is induced by the spontaneous mechanism or by the complex
parameters in the MSSM through radiative corrections, is determined
at zero and finite temperatures.
\end{titlepage}
\baselineskip=18pt
\setcounter{page}{2}
\setcounter{footnote}{0}
%
%
\section{Introduction}
The scenario of electroweak baryogenesis\cite{reviewEB}\  
requires that the electroweak phase transition (EWPT) to be of first 
order and that $CP$ violation is effective at that transition 
temperature. 
The EWPT is of first order only with too light Higgs boson in the  
minimal standard model\cite{BKS,EWPT-MSM}, and it is argued that
$CP$ violation in the CKM matrix cannot generate sufficient baryon 
asymmetry.
Both the requirements, however, will be fulfilled by some extension 
of the  standard model.
The minimal supersymmetric standard model (MSSM) is one of
promising candidates. 
When some of the scalar partners of the quarks and leptons are light 
enough, the EWPT becomes such a strongly first-order phase transition 
that the sphaleron process decouples just after it with acceptable 
mass of the lightest Higgs scalar\cite{light-stop1}.
Although the masses of the Higgs bosons in the MSSM are constrained by
some tree-level relations, they receive large radiative 
corrections from the top quark and squarks\cite{higgsmass-mssm1}.
This may broaden a window for successful baryogenesis by the MSSM.\par
The MSSM has many sources of $CP$ violations, in addition to the KM 
phase, such as the relative phases 
of the complex parameters $\mu$, the gaugino mass parameters and scalar
trilinear couplings, which are effective to generate the baryon 
asymmetry\cite{EWBsusy}.\  
Besides these complex phases, the relative phase of the expectation 
values of the two Higgs doublets could induce the source of baryon 
number\cite{higgs-phase}. 
This phase might be induced by radiative and finite-temperature effects
near the transition temperature\cite{spontCPV}\  or dynamically generated
near the bubble wall created at the EWPT, which we call 
{\it transitional $CP$ violation}\cite{FKOTT}.
This mechanism has been examined dynamically by solving equations of 
motion for the classical Higgs fields connecting the broken and 
symmetric phases. Then the potential for the fields are given by the 
effective potential at the transition temperature, which are 
approximated by a gauge-invariant polynomial whose coefficients are 
given by the effective parameters at that temperature\cite{FKOTmssm}.
It was shown that the contributions from charginos and neutralinos 
are important to trigger the $CP$ violation in the intermediate region.
These analyses contain undetermined parameters such as the transition 
temperature, thickness and velocity of the bubble wall, expectation 
values of the Higgs fields and the magnitude of explicit $CP$ violation
at the transition temperature.
These quantities should be determined by the parameters in the MSSM.\par
Now one of our main concerns is whether the EWPT in the MSSM is of 
first order strong enough for the sphaleron process to decouple after it,
with acceptable masses of Higgs bosons.
Recent study of the two-loop resummed effective potential at finite 
temperature suggests that the EWPT is strong enough for
$2\ltsim\tan\b\ltsim4$, $m_A\gtsim120\mbox{GeV}$ and a light Higgs 
with $m_h\ltsim85\mbox{GeV}$\cite{ewpt-2loop-mssm}. 
This and the previous analyses based on the one-loop resummed potential 
did not include the contributions from the charginos and neutralinos.
For the parameters which admit the strongly first-order EWPT,
we must examine whether efficient $CP$ violation exists at the transition
temperature. 
The effective potentials used in the previous study of the phase transition
were functions of only the $CP$-conserving order parameters, so that
they could not evaluate $CP$ violation in the Higgs sector at the 
EWPT.\par
In this paper, we use the effective potential in the MSSM, which includes
one-loop corrections from the top quark, top squarks, gauge bosons, 
charginos and  neutralinos, to evaluate the masses of the neutral 
Higgs bosons and to examine strength of the EWPT.
The masses are approximated by the eigenvalues of the matrix whose
elements are given by the second derivatives of the effective 
potential evaluated at the vacuum at zero temperature.
Although the mass formulas for the neutral Higgs bosons have been found
by fully-contained one-loop calculations\cite {higgsmass-mssm2} and by
two-loop calculations\cite{higgsmass-mssm3}, we adopt this method for
self-containedness. 
Without $CP$ violation, the minimum of the effective potential is 
parameterized by by the absolute value and the ratio of the VEVs of 
the two Higgs doublets. Extending the effective potential to include
the $CP$-violating order parameter and employing a numerical method to 
search a minimum of it, we find the magnitude of induced $CP$ violation
through radiative corrections from the superparticles when some of the 
parameters are complex-valued.
To find the transition temperature $T_C$ and the magnitude of the VEVs
of the Higgs fields at $T_C$, we numerically calculate the effective 
potential without use of the high-temperature expansion\cite{DJ},
and use the same numerical method as in zero-temperature case to search
a minimum of the effective potential. 
$CP$ violation in the Higgs sector will provide a boundary condition for
the equations which dynamically determines the profile of the bubble 
wall\cite{explicitCPV}.\par
This paper is organized as follows. In Section~2, we derive formulas 
for the neutral Higgs boson masses in the MSSM in the absence of $CP$ 
violation and a numerical method is introduced which can be applied to 
the case with $CP$ violation.
The method is extended to the effective potential at finite temperature
in Section~3.
Numerical results on the masses of the Higgs scalars and  transition 
temperature and strength of the EWPT are summarized in Section~4.
Section~5 is dedicated to concluding remarks. 
The formulas for the derivatives of the effective potential are 
summarized in the Appendix.
\section{Higgs Boson Masses}
If we write the VEVs of the two Higgs doubles as $\varphi_d$ and 
$\varphi_u$, the tree-level potential is given by
\begin{eqnarray}
 V_0 &=&
 m_1^2\varphi_d^\dagger\varphi_d+m_2^2\varphi_u^\dagger\varphi_u +
 (\mu B\varphi_u\varphi_d+\mbox{h.c})   \nonumber\\
 & &\qquad +
 {{g_2^2+g_1^2}\over8}
 (\varphi_d^\dagger\varphi_d-\varphi_u^\dagger\varphi_u)^2 +
 {{g_2^2}\over2}(\varphi_d^\dagger\varphi_d)(\varphi_u^\dagger\varphi_u),
                                 \label{eq:V0-1}
\end{eqnarray}
where we take $m_3^2\equiv \mu B$ to be real and positive by phase 
convention of the fields. Now we parameterize the VEVs 
as\footnote{The order parameter $v_4$ can be eliminated by the gauge 
transformation. So we shall set $v_4=0$ except when we numerically calculate
the Higgs boson masses in the presence of $CP$ violation.}
\begin{equation}
 \varphi_d={1\over{\sqrt2}}\pmatrix{v_1+iv_4\cr 0},\qquad
 \varphi_u={1\over{\sqrt2}}\pmatrix{0\cr v_2+iv_3}.
\end{equation}
Then the tree-level potential is expressed as
\begin{equation}
 V_0 = 
 \half m_1^2(v_1^2+v_4^2)+\half m_2^2(v_2^2+v_3^2) -
 m_3^2(v_1v_2-v_3v_4) +
 {{g_2^2+g_1^2}\over{32}}(v_1^2+v_4^2-v_2^2-v_3^2)^2.
\end{equation}
The minimum of this potential is given by
\begin{equation}
 v_1=v_0\cos\b,\quad v_2=v_0\sin\b,\quad v_3=v_4=0,
\end{equation}
with
\begin{equation}
 m_1^2 = m_3^2\tan\b-\half m_Z^2\cos(2\b), \quad
 m_2^2 = m_3^2\cot\b+\half m_Z^2\cos(2\b),
\end{equation}
so that $CP$ is conserved.
The effective potential at the one-loop level is given by
\begin{equation}
 V_{\rm eff}(\bm{v})=V_0(\bm{v})+\Delta V(\bm{v}),
                                      \label{eq:def-Veff-0}
\end{equation}
where
\begin{equation}
 \Delta V(\bm{v}) =
 \Delta_g V(\bm{v})+\Delta_t V(\bm{v})+\Delta_\st V(\bm{v})+
 \Delta_{\chi^\pm}V(\bm{v})+\Delta_{\chi^0}V(\bm{v})
\end{equation}
is the sum of the one-loop corrections.
Each term is given as follows.
\begin{eqnarray}
 \Delta_g V({\mib v}) &=&
 3\cdot2 F\left( m_W^2({\mib v})\right)+3F\left( m_Z^2({\mib v})\right),
          \label{eq:deltaV-gauge}\\
 \Delta_t V({\mib v}) &=&
  -4\cdot3\cdot F\bigl(m_t^2({\mib v})\bigr),\label{eq:deltaV-top}\\
 \Delta_\st V({\mib v}) &=&
 2\cdot3\cdot\sum_{a=1,2} F\left(m_{\st_a}^2({\mib v})\right),
       \label{eq:deltaV-stop}\\
 \Delta_{\chi^\pm}V({\mib v}) &=&
 -4\sum_{a=1,2} F\left(m_{\chi^\pm_a}^2({\mib v})\right),
       \label{eq:deltaV-chargino}\\
 \Delta_{\chi^0}V({\mib v}) &=&
 -2\sum_{a=1,2,3,4} F\left(m_{\chi^0_a}^2({\mib v})\right),
       \label{eq:deltaV-neutralino}
\end{eqnarray}
where
\begin{equation}
 F(m^2)\equiv
 {{m^4}\over{64\pi^2}}\left( \log{{m^2}\over{M_{\rm ren}^2}}
                             -{3\over2}\right), \label{eq:def-F}
\end{equation}
which was renormalized by the $\overline{\mbox{DR}}$-scheme.
The renormalization scale $M_{\rm ren}$ will be taken to be the weak 
scale. The masses-squared in (\ref{eq:deltaV-gauge}) and
(\ref{eq:deltaV-top}) are given by
\begin{eqnarray}
 m_W^2({\mib v})&=&{{g_2^2}\over4}(v_1^2+v_2^2+v_3^2+v_4^2),\qquad
 m_Z^2({\mib v})={{g_2^2+g_1^2}\over4}(v_1^2+v_2^2+v_3^2+v_4^2),\nonumber\\
 m_t^2({\mib v})&=&{{y_t^2}\over2}(v_2^2+v_3^2).
\end{eqnarray}
$m_{\st_a}^2({\mib v})$ in (\ref {eq:deltaV-stop}) are the eigenvalues of
the matrix
\begin{equation}
 M_\st^2 = \pmatrix{
  m_{\st_L}^2+m_t^2({\mib v})+{{3g_2^2-g_1^2}\over{12}}(v_1^2-v_2^2-v_3^2+v_4^2) &
  {{y_t}\over{\sqrt2}}\left[\mu(v_1+iv_4)+A_t(v_2-iv_3)\right] \cr
  {{y_t}\over{\sqrt2}}\left[\mu(v_1-iv_4)+A_t(v_2+iv_3)\right] &
  m_{\st_R}^2+m_t^2({\mib v})+{{g_1^2}\over6}(v_1^2-v_2^2-v_3^2+v_4^2) }.
     \label{eq:stop-mass-matrix}
\end{equation}
$m_{\chi^\pm_a}^2({\mib v})$ in (\ref{eq:deltaV-chargino}) are the 
eigenvalues of the matrix $M_{\chi^\pm}^\dagger M_{\chi^\pm}$ with
\begin{equation}
 M_{\chi^\pm} =
 \pmatrix{ M_2 & -{i\over{\sqrt2}}g_2(v_2-iv_3) \cr
           -{i\over{\sqrt2}}g_2(v_1-iv_4) & -\mu },
             \label{eq:chargino-mass-matrix}
\end{equation}
and $m_{\chi^0_a}^2({\mib v})$ in (\ref {eq:deltaV-neutralino}) are the
eigenvalues of the matrix $M_{\chi^0}^\dagger M_{\chi^0}$ with
\begin{equation}
 M_{\chi^0} =
 \pmatrix{ M_2 & 0 & -{i\over2}g_2(v_1-iv_4) &  {i\over2}g_2(v_2-iv_3) \cr
           0 & M_1 &  {i\over2}g_1(v_1-iv_4) & -{i\over2}g_1(v_2-iv_3) \cr
         -{i\over2}g_2v_1(v_1-iv_4) & {i\over2}g_1(v_1-iv_4) & 0 & \mu \cr
   {i\over2}g_2(v_2-iv_3) & -{i\over2}g_1(v_2-iv_3) & \mu & 0 }.
          \label{eq:neutralino-mass-matrix}
\end{equation}
In general, $\mu$, $A_t$, $M_2$ and $M_1$ are complex-valued but
we assume them to be real until we discuss $CP$ violation.\par
The minimization conditions of the effective potential relate the 
mass parameters in the Higgs potential to the VEVs of the Higgs fields:
\begin{eqnarray}
 m_1^2 &=&
 m_3^2\tan\b-\half m_Z^2\cos(2\b)-
 {1\over{v_1}}{{\del\Delta V({\mib v})}\over{\del v_1}},\nonumber\\
 m_2^2 &=&
 m_3^2\cot\b+\half m_Z^2\cos(2\b)-
 {1\over{v_2}}{{\del\Delta V({\mib v})}\over{\del v_2}},
                \label{eq:m1-m2}
\end{eqnarray}
which are used to eliminate $m_1^2$ and $m_2^2$ in favor of 
$v_0\equiv\absv{\bm{v}}$ and $\tan\b=v_2/v_1$. These are equivalent to the
tadpole conditions\cite{higgsmass-mssm2}.
The masses of the $CP$-even Higgs bosons are the eigenvalues of the 
matrix
\begin{equation}
 {\cal M}_h^2 \equiv
 \pmatrix{
  {{\del^2V_{\rm eff}({\mib v})}\over{\del v_1^2}} &
  {{\del^2V_{\rm eff}({\mib v})}\over{\del v_1\del v_2}} \cr
  {{\del^2V_{\rm eff}({\mib v})}\over{\del v_1\del v_2}} &
  {{\del^2V_{\rm eff}({\mib v})}\over{\del v_2^2}} },
        \label{eq:def-Mh-matrix}
\end{equation}
where the derivatives should be evaluated at the vacuum.
The mass of the $CP$-odd scalar is given by
\begin{equation}
  m_A^2 = {1\over{\cos^2\b}}
  {{\del^2V_{\rm eff}({\mib v})}\over{\del v_3^2}}.\label{eq:def-mA}
\end{equation}
By use of (\ref{eq:m1-m2}), the second derivatives evaluated at the vacuum
are reduced to
\begin{eqnarray}
 {{\del^2V_{\rm eff}({\mib v})}\over{\del v_1^2}} &=&
 m_3^2\tan\b + m_Z^2\cos^2\b +
 v_1{\del\over{\del v_1}}\left(
 {1\over{v_1}}{{\del\Delta V({\mib v})}\over{\del v_1}}\right), 
                                \label{eq:d11-Veff-0}\\
 {{\del^2V_{\rm eff}({\mib v})}\over{\del v_2^2}} &=&
 m_3^2\cot\b + m_Z^2\sin^2\b +
 v_2{\del\over{\del v_2}}\left(
 {1\over{v_2}}{{\del\Delta V({\mib v})}\over{\del v_2}}\right),
                                \label{eq:d22-Veff-0}\\
 {{\del^2V_{\rm eff}({\mib v})}\over{\del v_1\del v_2}} &=&
 -m_3^2 - m_Z^2\sin\b\cos\b +
 {{\del^2\Delta V({\mib v})}\over{\del v_1\del v_2}},
                                           \label{eq:d12-Veff-0}\\
 {{\del^2V_{\rm eff}({\mib v})}\over{\del v_3^2}} &=&
 m_3^2\cot\b - {1\over{v_2}}{{\del\Delta V({\mib v})}\over{\del v_2}}+
 {{\del^2\Delta V({\mib v})}\over{\del v_3^2}}.  \label{eq:d33-Veff-0}
\end{eqnarray}
The expressions of the derivatives of $\Delta V({\mib v})$ are 
summarized in the Appendix.\par
In addition to the evaluation of the masses by use of these formulas,
we adopt a fully numerical method. In this method, the effective potential
defined by (\ref{eq:def-Veff-0}) is calculated at every
$\bm{v}=(v_1,v_2,v_3,v_4)$, where the mass eigenvalues are evaluated
numerically. 
For a given set of $(v_0,\tan\b)$ and $m_3^2$, the mass parameters in
the Higgs potential are determined by (\ref{eq:m1-m2}).
The minimum of the effective potential is searched 
by use of the downhill simplex algorithm\cite{num-recipe}\  
starting from a randomly generated simplex in the restricted space of 
$(v_1,v_2,v_3)$ with $v_4=0$. 
Once the minimum is found, the second derivatives of the effective potential
with respect to $(v_1,v_2,v_3,v_4)$ are numerically evaluated. 
We have checked, in the absence of $CP$ violation,
that the minimum coincides with the prescribed $(v_0,\tan\b)$ and
that the four-by-four matrix of the second derivatives is completely 
divided into to the two sectors of $CP$ eigenmodes and the eigenvalues
coincides with those obtained by use of the formulas above.
This numerical method can be applied to the case with $CP$ violation.
In the presence of $CP$ violation such as the relative phases of 
complex parameters, $CP$-violating order parameter $v_3$ is induced
and the $CP$ eigenstates of the Higgs sector mix to make the mass
eigenstates.
\section{Finite-Temperature Effective Potential}
At finite temperatures, the one-loop corrections in 
(\ref{eq:deltaV-gauge})--(\ref{eq:deltaV-neutralino}) are modified to 
include the finite-temperature effects:
\begin{eqnarray}
 \Delta_g V({\mib v};T) &=& \Delta_g V({\mib v})+
 {{T^4}\over{2\pi^2}}\left[
     6I_B\left({{m_W^2({\mib v})}\over{T^2}}\right) +
     3I_B\left({{m_Z^2({\mib v})}\over{T^2}}\right)\right],
          \label{eq:deltaV-gauge-T}\\
 \Delta_t V({\mib v};T) &=& \Delta_t V({\mib v};T) -
 12\cdot {{T^4}\over{2\pi^2}}
    I_F\left({{m_t^2({\mib v})}\over{T^2}}\right),\label{eq:deltaV-top-T}\\
 \Delta_\st V({\mib v};T) &=& \Delta_\st V({\mib v}) +
 6\cdot {{T^4}\over{2\pi^2}}
   \sum_{a=1,2}I_B\left({{m_{\st_a}^2({\mib v})}\over{T^2}}\right),
       \label{eq:deltaV-stop-T}\\
 \Delta_{\chi^\pm}V({\mib v};T) &=& \Delta_{\chi^\pm}V({\mib v})
 -4\cdot{{T^4}\over{2\pi^2}}
   \sum_{a=1,2} I_F\left({{m_{\chi^\pm_a}^2({\mib v})}\over{T^2}}\right),
       \label{eq:deltaV-chargino-T}\\
 \Delta_{\chi^0}V({\mib v};T) &=& \Delta_{\chi^0}V({\mib v})
 -2\cdot{{T^4}\over{2\pi^2}}\sum_{a=1,2,3,4} 
   I_F\left({{m_{\chi^0_a}^2({\mib v})}\over{T^2}}\right),
       \label{eq:deltaV-neutralino-T}
\end{eqnarray}
where the functions $I_B(a^2)$ and $I_F(a^2)$ are defined by
\begin{equation}
 I_{B,F}(a^2) = \int_0^\infty dx\; x^2
 \log\left(1\mp e^{-\sqrt{x^2+a^2}}\right).\label{eq:def-Ib-If}
\end{equation}
The effective potential at finite temperature is calculated at each
${\mib v}$ by numerically evaluating the mass-squared eigenvalues and
inserting them into the expressions 
(\ref{eq:deltaV-gauge-T})--(\ref{eq:deltaV-neutralino-T}). The 
integrals defined in (\ref{eq:def-Ib-If}) are numerically calculated 
without use of the high-temperature expansions\cite{DJ}.\par
For a given set of parameters, the minimum of the effective potential
is searched at various temperatures by the method stated in the previous
section. Near the transition temperature, several numbers of 
starting simplexes are generated and the minimum reached starting from
each simplex is found. 
The temperature at which two degenerate minima are found is 
defined to be the transition temperature $T_C$ of the first-order EWPT.
Then we examine whether the condition is satisfied for the sphaleron 
process to decouple just after the EWPT\cite{sph-decouple}
\begin{equation}
 {{v_C}\over{T_C}} = 
 \lim_{T\uparrow T_C}{{\absv{{\mib v}(T)}}\over{T}} > 1.
       \label{eq:sphaleron-decouple}
\end{equation}
If $v_C=0$, the EWPT is of second order.
We executed this minimum search for various sets of parameters to find
the order of the EWPT and $T_C$, and measured $v_C$ and $\tan\b$ at 
$T_C$ when the EWPT is of first order.
\section{Numerical Results}
\subsection{$\mib{CP}$-conserving case}
Among many parameters in the MSSM, the mass parameters $m_1^2$ and 
$m_2^2$ in the Higgs potential are determined by (\ref{eq:m1-m2})
in the absence of $CP$ violation.
Throughout this paper, we take $v_0=246\mbox{GeV}$, 
$m_W=80.3\mbox{GeV}$, $m_Z=91.2\mbox{GeV}$ and $m_t=175\mbox{GeV}$.
The rest of the parameters are $m_3^2$, $\tan\b$, $\mu$, $A_t$,
$m_{\st_L}$, $m_{\st_R}$, $M_2$ and $M_1$.
For definiteness, we take $M_2=M_1$, $m_{\st_L}=400\mbox{GeV}$ and
$A_t=10\mbox{GeV}$.\par
Before presenting the numerical results on the Higgs masses and $CP$ 
violation, we note that the contributions from the charginos and 
neutralinos are not negligible compared to those from the top quarks 
and squarks.
For example, consider the contributions to the first derivatives 
appearing in (\ref{eq:m1-m2}), which have the form of
${1\over{v_i}}{{\del m^2}\over{\del v_i}}F'(m^2)$.
The contributions from the gauge bosons is smaller than those 
from the top quark, since for $\tan\b=5$ 
\begin{equation}
 {1\over{v_2}}{{\del m^2_t({\mib v})}\over{\del v_2}} = 
 {{2m_t^2}\over{v_0^2\sin^2\b}}=y_t^2
 \simeq 1.0526,\qquad
 {1\over{v_i}}{{\del m_W^2({\mib v})}\over{\del v_i}} ={{g_2^2}\over2}
 \simeq 0.2131,
\end{equation}
which multiply $-12F'(m_t^2)$ and $6F'(m_W^2)$, respectively.  
For the case of the stop and charginos, these factors are replaced with
(\ref{eq:d2-m-stop}) and (\ref{eq:d1-m-chargino}) or (\ref{eq:d2-m-chargino}),
which depend not only on the couplings but also on $\mu$, $\tan\b$ and
the soft-SUSY-breaking masses. When $m_\st^2$ is the same order as
$m_{\chi^\pm}$, there is no reason for the stop contribution to become
much larger than the chargino contributions.
Now let us denote
\begin{eqnarray}
 \Delta_\st m_1^2 &=&
 -{1\over{v_1}}{{\del\Delta_\st V({\mib v})}\over{\del v_1}},\qquad
 \Delta_\st m_2^2 =
 -{1\over{v_2}}{{\del\Delta_\st V({\mib v})}\over{\del v_2}},\nonumber\\
 \Delta_{\chi^\pm} m_1^2 &=&
 -{1\over{v_1}}{{\del\Delta_{\chi^\pm}V({\mib v})}\over{\del v_1}},\qquad
 \Delta_{\chi^\pm} m_2^2 =
 -{1\over{v_2}}{{\del\Delta_{\chi^\pm}V({\mib v})}\over{\del v_2}},
\end{eqnarray}
and calculate them for $M_2=300\mbox{GeV}$. Several 
numerical values are presented in Table~\ref{tab:dV-stop-chargino}.
\begin{table}
\caption{Contributions to the equations relating the mass parameters 
in the Higgs potential to the VEVs of the Higgs fields from the stop 
and charginos. $(\tan\b,\mu)=(2,-20\mbox{GeV})$ and
$(\tan\b,\mu)=(5,-50\mbox{GeV})$ correspond to the zero stop mixing case.}
\label{tab:dV-stop-chargino}
\begin{center}
\small{
\begin{tabular}{|ccc|cccc|}
\hline
 $\tan\b$ & $\mu$(GeV)&$m_{\st_R}$(GeV)&
 $\Delta_\st m_1^2(\mbox{GeV}^2)$ & $\Delta_\st m_2^2(\mbox{GeV}^2)$ &
 $\Delta_{\chi^\pm} m_1^2(\mbox{GeV}^2)$ &
 $\Delta_{\chi^\pm} m_2^2(\mbox{GeV}^2)$ \\
\hline
 $2$ & $-20$  &  $0$  & $-1.323\times10^3$ & $-7.402\times10^3$ &
                  $5.723\times10^2$ & $6.967\times10^2$ \\
 $2$ & $-20$  & $300$ & $-1.596\times10^3$ & $-1.131\times10^4$ &
                  $5.723\times10^2$ & $6.967\times10^2$ \\
 $2$ & $-300$ &  $0$  & $-5.900\times10^3$ & $-7.482\times10^3$ &
                  $-4.799\times10^3$ & $1.263\times10^3$ \\
 $2$ & $-300$ & $300$ & $-7.089\times10^3$ & $-1.131\times10^4$ &
                  $-4.799\times10^3$ & $1.263\times10^3$ \\
 $5$ & $-50$  &  $0$  & $-1.302\times10^3$ & $-5.831\times10^3$ &
                  $8.050\times10^1$ & $7.293\times10^2$ \\
 $5$ & $-50$  & $300$ & $-1.573\times10^3$ & $-9.017\times10^3$ &
                  $8.050\times10^1$ & $7.293\times10^2$ \\
 $5$ & $-300$ &  $0$  & $-4.688\times10^3$ & $-5.824\times10^3$ &
                  $-3.769\times10^3$ & $1.628\times10^3$ \\
 $5$ & $-300$ & $300$ & $-5.640\times10^3$ & $-8.999\times10^3$ &
                  $-3.769\times10^3$ & $1.628\times10^3$ \\
\hline
\end{tabular}
}
\end{center}
\end{table}
This suggests that when $M_2\sim\absv{\mu}\sim m_{\st_L}$, 
$\Delta_\st m_1^2$ and $\Delta_\st m_2^2$ are of the same order as
$\Delta_{\chi^\pm} m_1^2$ and $\Delta_{\chi^\pm} m_2^2$.
As seen from (\ref{eq:def-d-R-stop}) and (\ref{eq:def-n-R-chargino}),
the factors multiplying $F'(m^2)$, which are the couplings squared in 
the case of the quarks and gauge bosons, are corrected by 
$n_\st^{(1)}/R_\st$,  $n_\chi^{(1)}/R_\chi$ and $n_\chi^{(2)}/R_\chi$
for the stops and charginos, respectively. For the case of $\tan\b=5$,
$\mu=-300\mbox{GeV}$ and $m_{\st_R}=0$ in Table~\ref{tab:dV-stop-chargino},
$n_\st^{(1)}/R_\st\simeq 0.53$ while
$n_\chi^{(1)}/R_\chi\simeq-13.4$ and $n_\chi^{(1)}/R_\chi\simeq 2.85$,
which are large enough to compensate the difference between the gauge 
and Yukawa coupling constants.
As for the neutralino, although its contributions
cannot be expressed in a compact form as (\ref{eq:di-deltaV-stop}),
it is natural to expect the neutralino contributions to be the same order
as the chargino, as far as $M_2\sim M_1$. For $\tan\b=5$ and
$\mu=-300\mbox{GeV}$, we have
\begin{equation}
 {1\over{v_1}}{{\del\Delta_{\chi^0}V({\mib v})}\over{\del v_1}} =
 2.374\times10^3\mbox{GeV}^2,\qquad
 {1\over{v_2}}{{\del\Delta_{\chi^0}V({\mib v})}\over{\del v_2}} =
 -1.013\times10^3\mbox{GeV}^2,
\end{equation}
which are the same order as the chargino contributions.
As for the second derivatives, we find that the contributions from
the charginos and neutralinos are the same order as those from the 
stops for the parameters we adopted in the numerical analyses.
We also observed that if we omit $\Delta_{\chi^\pm}V({\mib v})$
and $\Delta_{\chi^0}V({\mib v})$ to determine $m_1^2$ and $m_2^2$ in 
favor of $v_0$ and $\tan\b$ by (\ref{eq:m1-m2}), the numerical method 
to search the minimum results in a different point in $(v_1,v_2)$-plane
from $(v_0\cos\b,v_0\sin\b)$
with a deviation of about $5\%$ for $\tan\b>10$ and of about
$70\%$ for $\tan\b\ltsim2$.
Thus, as long as the soft mass parameters $M_2$, $M_1$ and
$m_{\st_{L,R}}$ are of the same order, the contributions from the 
charginos and neutralinos are comparable to those from the top squarks.\par
\begin{figure}
 \epsfxsize=15.0cm
 \centerline{\epsfbox{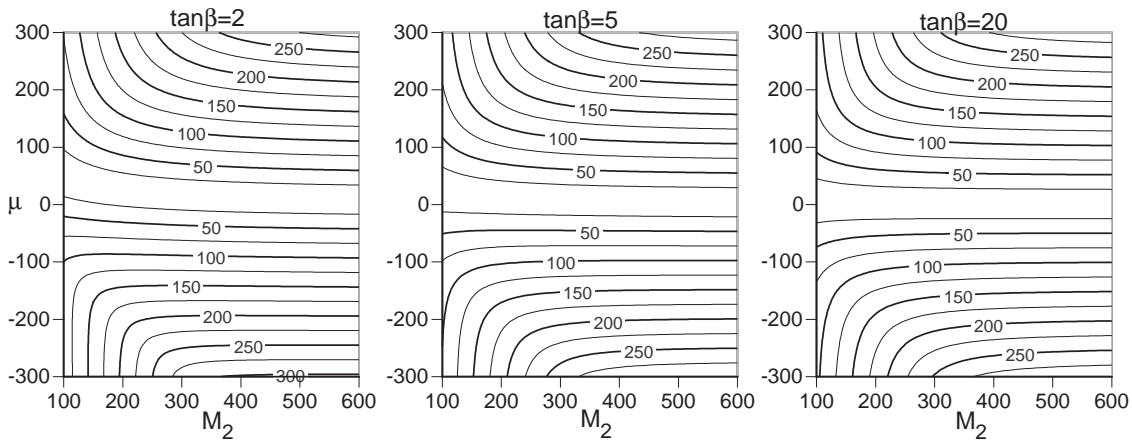}}
 \caption{Contour plots of the mass of the lighter chargino as 
 function of the MSSM parameters $M_2$ and $\mu$ for $\tan\b=2$,
 $5$ and $20$, respectively. All the mass units are GeV.}
 \label{fig:m_chargino}
\end{figure}

Now we show results on the masses of the neutral Higgs bosons.
We examine dependence of the mass of the lighter scalar $m_h$ on
the pseudoscalar mass $m_A$ and $M_2=M_1$. In practice, we calculate
$m_h$ and $m_A$ as functions of $(m_3^2, M_2)$ for a fixed set of
$(\tan\b,\mu,m_{\st_R})$ and make a contour plot of $m_h$ in
$(M_2,m_A)$-plane.
The mass of the lighter chargino $m_{\chi^\pm_1}$ is constrained to be
$m_{\chi^\pm_1}>65.7\mbox{GeV}$\cite{pd}, which restricts $\mu$ and 
$M_2$. According to the tree-level mass formula 
(\ref{eq:chargino-mass}), the mass is plotted as a function of
$M_2$ and $\mu$ in Fig.~\ref{fig:m_chargino}. This shows that the
lower limit is satisfied for the whole range of $M_2$ we studied, 
if we take $\absv{\mu}\gtsim100\mbox{GeV}$.
The limits on the masses of the lighter Higgs scalar and pseudoscalar
are now $m_h>62.5\mbox{GeV}$ and $m_A>62.5\mbox{GeV}$\cite{pd}, 
although more stringent bounds are reported 
$m_h\gtsim 75\mbox{GeV}$\cite{Barger}.
The results on $m_h$ are plotted in Figs.~\ref{fig:mh-(2,-300)} and
\ref{fig:mh-(2,300)} for $\tan\b=2$, in Figs.~\ref{fig:mh-(5,-300)} 
and \ref{fig:mh-(5,300)} and in Figs.~\ref{fig:mh-(20,-300)} and
\ref{fig:mh-(20,300)}. We also calculated $m_h$ for 
$\absv{\mu}=100\mbox{GeV}$ and $200\mbox{GeV}$ and found that
$m_h$ behaves in the same manner as the case of 
$\absv\mu=300\mbox{GeV}$ but its value becomes a bit smaller
for smaller $\absv\mu$.
\begin{figure}
 \epsfxsize=15.0cm
 \centerline{\epsfbox{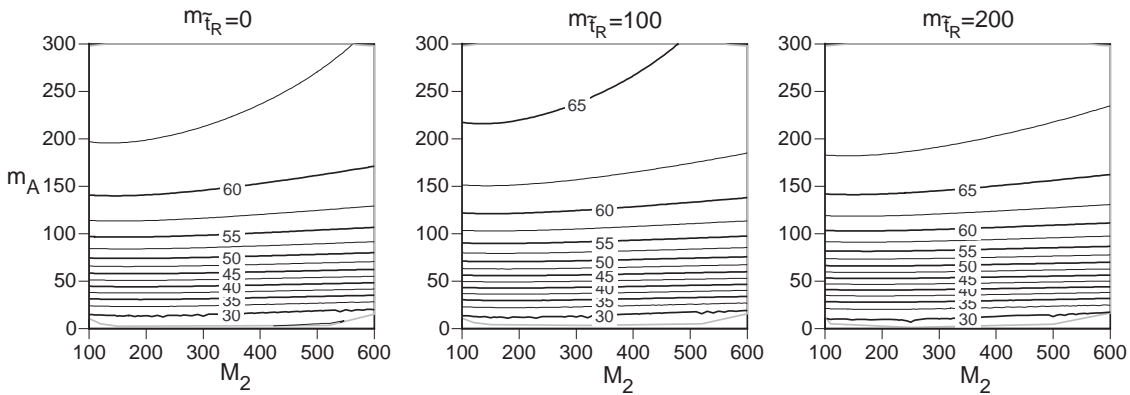}}
 \caption{Contour plots of $m_h$ as 
 function of the MSSM parameters $M_2$ and $m_A$ for 
 $\tan\b=2$, $\mu=-300\mbox{GeV}$ and $m_{\st_R}=0$,
 $100\mbox{GeV}$ and $200\mbox{GeV}$, respectively. 
 All the mass units are GeV.}
 \label{fig:mh-(2,-300)}
\end{figure}
\begin{figure}
 \epsfxsize=15.0cm
 \centerline{\epsfbox{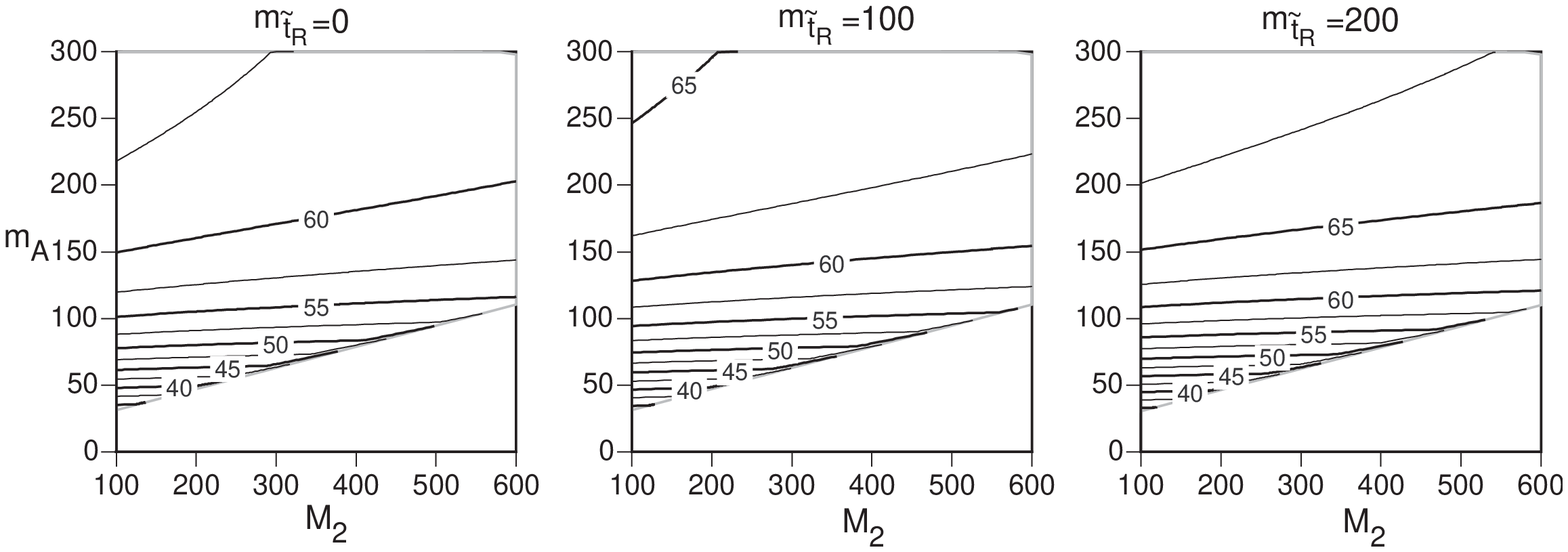}}
 \caption{The same as Fig.~\ref{fig:mh-(2,-300)} but for 
 $\tan\b=2$, $\mu=300\mbox{GeV}$ and $m_{\st_R}=0$,
 $100\mbox{GeV}$ and $200\mbox{GeV}$, respectively. 
 All the mass units are GeV.}
 \label{fig:mh-(2,300)}
\end{figure}
\begin{figure}
 \epsfxsize=15.0cm
 \centerline{\epsfbox{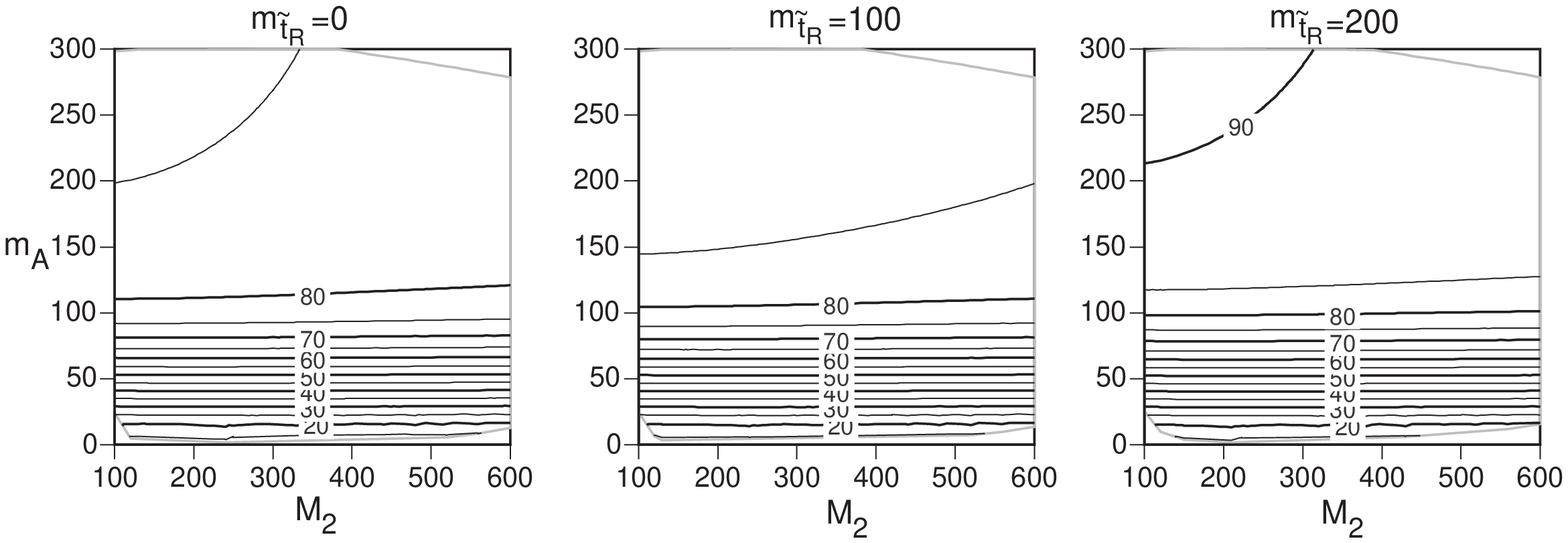}}
 \caption{The same as Fig.~\ref{fig:mh-(2,-300)} but for 
 $\tan\b=5$, $\mu=-300\mbox{GeV}$ and $m_{\st_R}=0$,
 $100\mbox{GeV}$ and $200\mbox{GeV}$, respectively. 
 All the mass units are GeV.}
 \label{fig:mh-(5,-300)}
\end{figure}
\begin{figure}
 \epsfxsize=15.0cm
 \centerline{\epsfbox{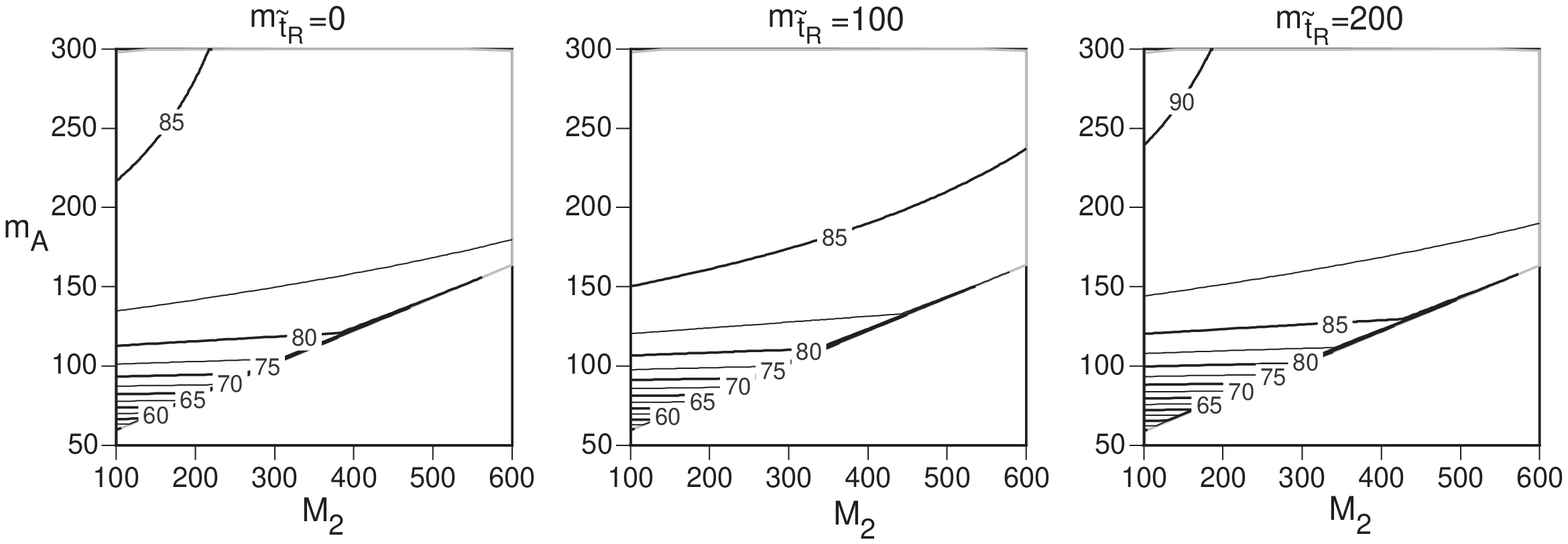}}
 \caption{The same as Fig.~\ref{fig:mh-(2,-300)} but for 
 $\tan\b=5$, $\mu=300\mbox{GeV}$ and $m_{\st_R}=0$,
 $100\mbox{GeV}$ and $200\mbox{GeV}$, respectively. 
 All the mass units are GeV.}
 \label{fig:mh-(5,300)}
\end{figure}
\begin{figure}
 \epsfxsize=15.0cm
 \centerline{\epsfbox{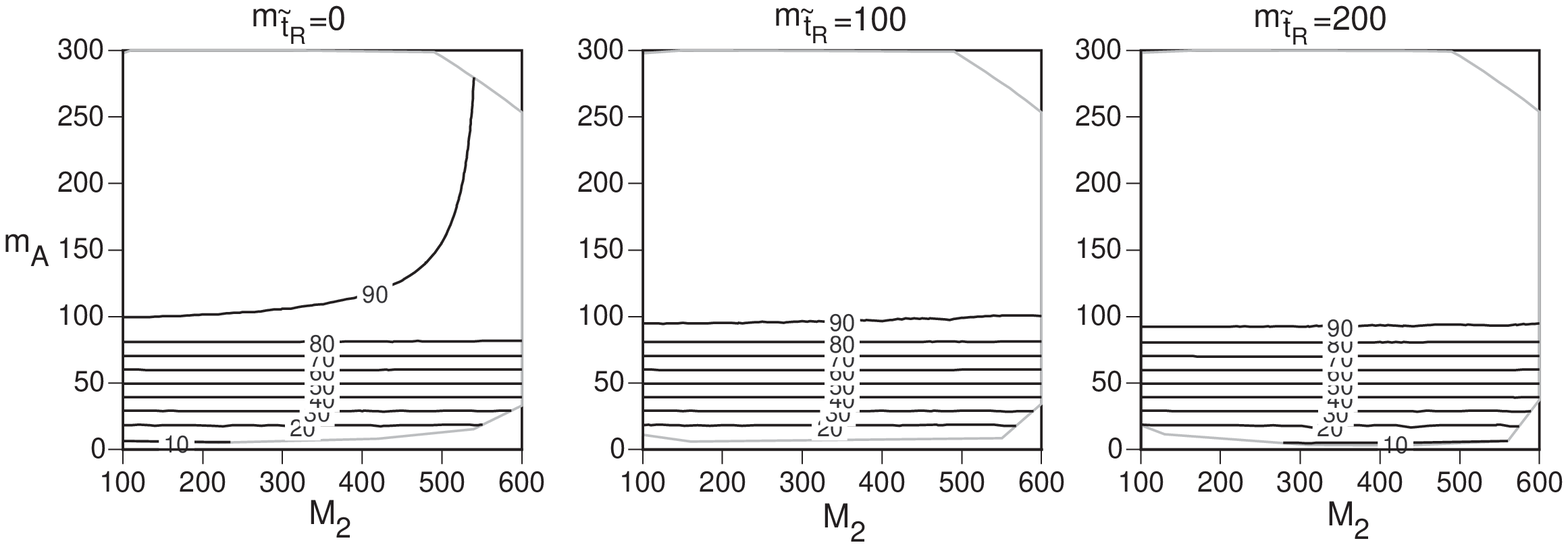}}
 \caption{The same as Fig.~\ref{fig:mh-(2,-300)} but for 
 $\tan\b=20$, $\mu=-300\mbox{GeV}$ and $m_{\st_R}=0$,
 $100\mbox{GeV}$ and $200\mbox{GeV}$, respectively. 
 All the mass units are GeV.}
 \label{fig:mh-(20,-300)}
\end{figure}
\begin{figure}
 \epsfxsize=15.0cm
 \centerline{\epsfbox{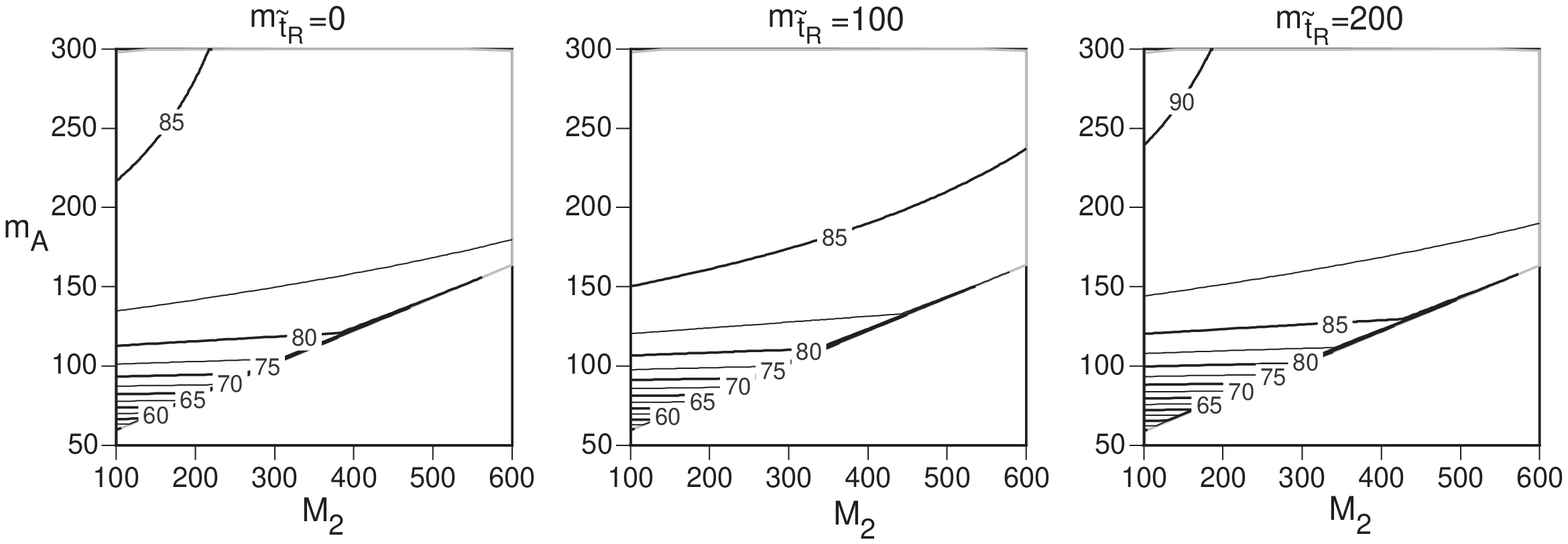}}
 \caption{The same as Fig.~\ref{fig:mh-(2,-300)} but for 
 $\tan\b=20$, $\mu=300\mbox{GeV}$ and $m_{\st_R}=0$,
 $100\mbox{GeV}$ and $200\mbox{GeV}$, respectively. 
 All the mass units are GeV.}
 \label{fig:mh-(20,300)}
\end{figure}
For $\mu=300\mbox{GeV}$, in the blank region at small $m_A$ and large
$M_2$ region, the point ${\mib v}=(v_0\cos\b,v_0\sin\b)$ is not
a local minimum but a saddle point. This region is broader for larger
$\tan\b$, which corresponds to smaller top Yukawa coupling. This is 
because for larger $M_2$, the contributions from the charginos and 
neutralinos to the effective potential, which are negative, 
dominate over the bosonic contributions and make the vacuum unstable. 
For $\tan\b=2$, the experimental bound on $m_h$ is satisfied for
$m_A\gtsim200-300\mbox{GeV}$, depending on $\mu$ and $M_2$.
For $\tan\b\ge5$, it is satisfied for $m_A\gtsim100\mbox{GeV}$.\par
At finite temperatures, the minimum of the effective potential differs 
from that at zero temperature. What we concern here are the order of 
the EWPT and transition temperature $T_C$, at which two minima of the 
effective potential degenerate, and the location of the minimum at $T_C$
when it is of first order. 
These are important ingredients for the electroweak baryogenesis.
For definiteness, we take $\mu=-300\mbox{GeV}$, $M_2=M_1=350\mbox{GeV}$, 
$m_{\st_L}=400\mbox{GeV}$ and $A_t=10\mbox{GeV}$.
By use of the numerical method explained in the previous section, we
search the minimum of the effective potential at various temperatures
to find the transition temperature.
This analysis is done for $\tan\b=2$, $5$ and $20$ with various 
$m_{\st_R}\ge0$ and two values of $m_3^2$ being tuned so that 
$m_h\simeq 62.5\mbox{GeV}$ and $m_h\simeq 80\mbox{GeV}$, respectively,
for $m_{\st_R}=0$.\footnote{For $\tan\b=2$ 
and the rest of the parameters given above, $m_h$ cannot be so large as
$70\mbox{GeV}$. So the results are presented only for
$m_h\simeq 62.5\mbox{GeV}$.}\par
\begin{figure}
 \epsfysize=4.5cm
 \centerline{\epsfbox{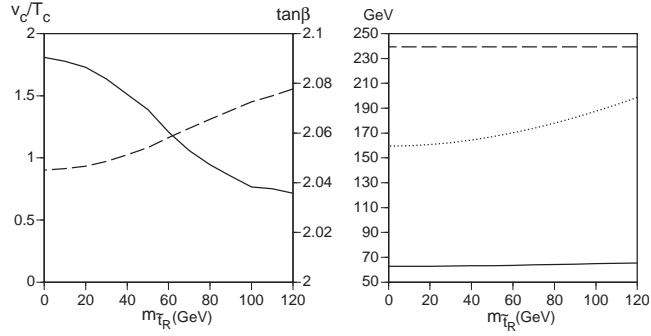}}
 \caption{Dependence of $v_C/T_C$ (solid curve) and $\tan\b(T_C)$ 
 (dashed curve) on $m_{\st_R}$ for $m_3^2=2.5\times10^4\mbox{GeV}^2$.
 $m_h$ (solid curve), $m_A$ (dashed curve) and $m_{\st_1}$ (dotted 
 curve) are also shown for the same parameter set.}
 \label{fig:Tc-tanb2}
\end{figure}

For $\tan\b=2$ and small $m_{\st_R}$, it is difficult to have
$m_h$ larger than $70\mbox{GeV}$ as seen from Fig.~\ref{fig:mh-(2,-300)}.
We have adopted $m_3^2=2.5\times10^4\mbox{GeV}^2$, for which
$m_h\simeq 62.5\mbox{GeV}$ when $m_{\st_R}=0$.
Dependences of $v_C/T_C$, $\tan\b(T_C)$ and the masses on $m_{\st_R}$ 
are plotted in Fig.~\ref{fig:Tc-tanb2}.
The condition for the sphaleron decoupling after the EWPT 
(\ref{eq:sphaleron-decouple}) is satisfied for 
$m_{\st_R}\ltsim75\mbox{GeV}$, for which $m_h\ltsim64\mbox{GeV}$ and
$m_A\simeq 239\mbox{GeV}$. 
The transition temperature varies from $T_C=77.3\mbox{GeV}$ 
($m_{\st_R}=0$) to $88.7\mbox{GeV}$ ($m_{\st_R}=120\mbox{GeV}$)
monotonously.
$\tan\b$ at $T_C$ is almost independent of $m_{\st_R}$ and remains
to be the zero temperature value.\par
\begin{figure}
 \epsfysize=4.5cm
 \centerline{\epsfbox{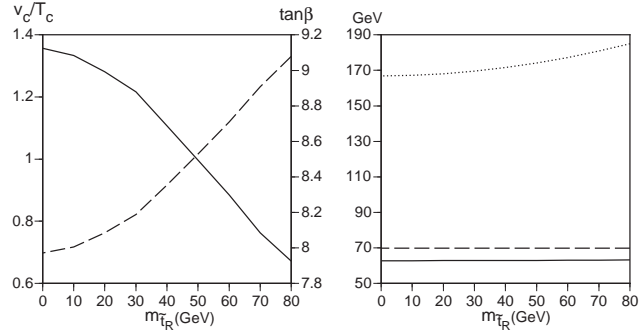}}
 \caption{The same as Fig.~\ref{fig:Tc-tanb2} but for $\tan\b=5$ and
 $m_3^2=3050\mbox{GeV}^2$.}
 \label{fig:Tc-tanb5-1}
\end{figure}

\begin{figure}
 \epsfysize=4.5cm
 \centerline{\epsfbox{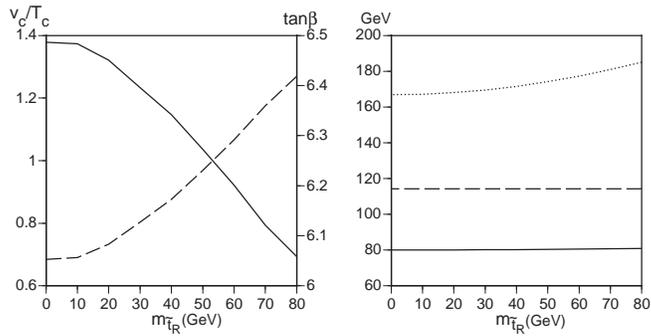}}
 \caption{The same as Fig.~\ref{fig:Tc-tanb2} but for $\tan\b=5$ and
 $m_3^2=4624\mbox{GeV}^2$.}
 \label{fig:Tc-tanb5-2}
\end{figure}

%
%
For $\tan\b=5$, we have taken $m_3^2=3050\mbox{GeV}^2$ and 
$4624\mbox{GeV}^2$, which correspond to $m_h\simeq 62.5\mbox{GeV}$ and
$80\mbox{GeV}$, respectively. $T_C$ is monotonously decreasing with 
respect to $m_{\st_R}$ and for the former case, 
$93.2\mbox{GeV}\le T_C\le 100.5\mbox{GeV}$, while 
$93.0\mbox{GeV}\le T_C\le 100.2\mbox{GeV}$ for the latter case.
Dependence of $T_C$ on $m_3^2$, so that on $m_A$, appears to be 
weak. $v_C/T_C$, $\tan\b(T_C)$ and the masses are plotted in 
Fig.~\ref{fig:Tc-tanb5-1} and Fig.~\ref{fig:Tc-tanb5-2}.
The condition (\ref{eq:sphaleron-decouple}) is satisfied for
$m_{\st_R}\ltsim50\mbox{GeV}$. $\tan\b$ at $T_C$ receives 
finite-temperature corrections to become about $20$\% larger than the
zero-temperature value for the case of the larger $m_A$.
\par
\begin{figure}
 \epsfysize=4.5cm
 \centerline{\epsfbox{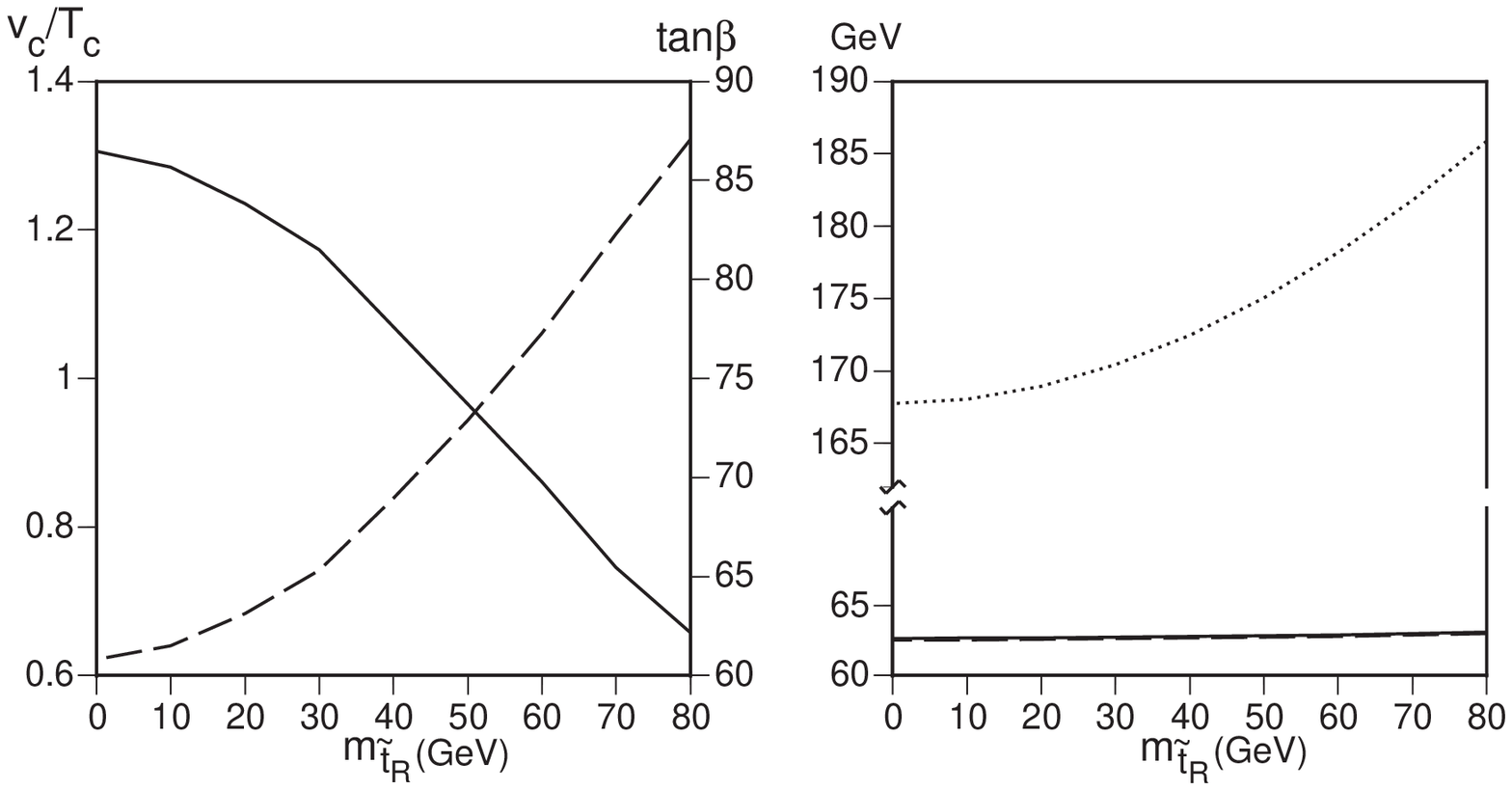}}
 \caption{The same as Fig.~\ref{fig:Tc-tanb2} but for $\tan\b=20$ and
 $m_3^2=2308\mbox{GeV}^2$.}
 \label{fig:Tc-tanb20-1}
\end{figure}

\begin{figure}
 \epsfysize=4.5cm
 \centerline{\epsfbox{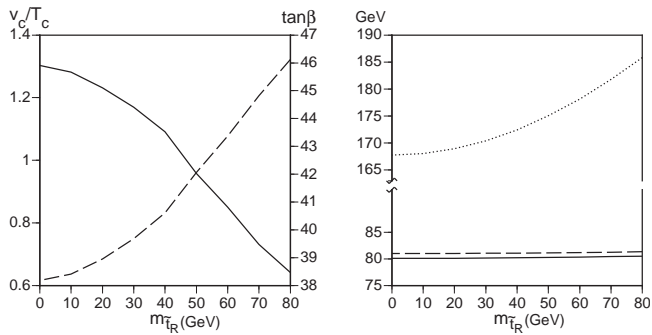}}
 \caption{The same as Fig.~\ref{fig:Tc-tanb2} but for $\tan\b=20$ and
 $m_3^2=2440\mbox{GeV}^2$.}
 \label{fig:Tc-tanb20-2}
\end{figure}

For $\tan\b=20$, we adopted $m_3^2=2308\mbox{GeV}^2$ and $2440\mbox{GeV}^2$.
Dependence of $T_C$ on $m_{\st_R}$ and $m_3^2$ is similar to the 
previous examples of $\tan\b=5$. 
Now $97.2\mbox{GeV}\le T_C\le 104.5\mbox{GeV}$ for both choices of $m_3^2$.
$v_C/T_C$, $\tan\b(T_C)$ and the masses are plotted in 
Fig.~\ref{fig:Tc-tanb20-1} and Fig.~\ref{fig:Tc-tanb20-2}.
In this case, $\tan\b(T_C)$ drastically deviate from the 
zero-temperature value.\par
\begin{figure}
 \epsfysize=6.5cm
 \centerline{\epsfbox{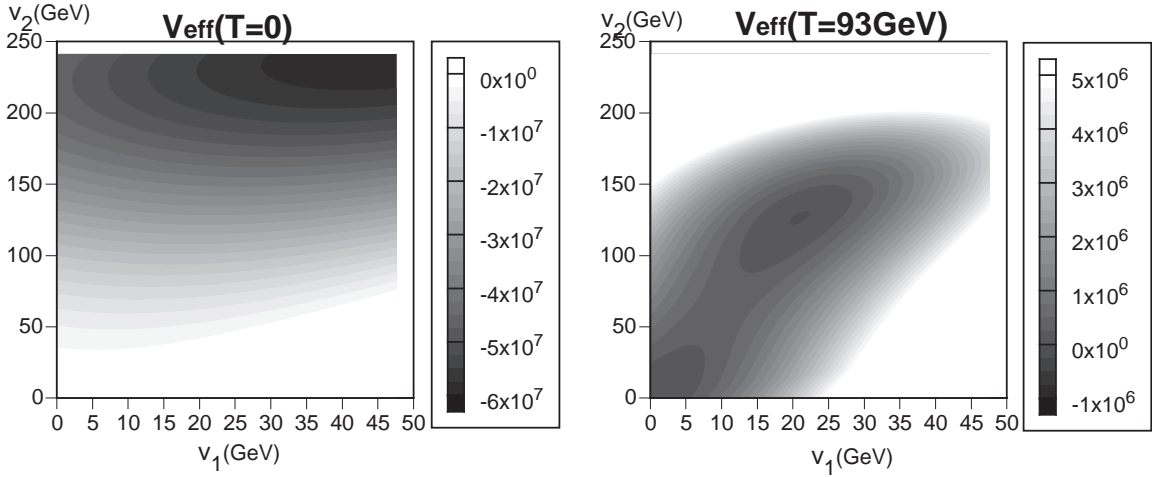}}
 \caption{Contour plots of the effective potential at $T=0$ and $T=T_C$ 
 for $\tan\b=5$, $m_3^2=4624\mbox{GeV}^2$ and $m_{\st_R}=0$.}
 \label{fig:Veff-contour}
\end{figure}

In order to determine the profile of the bubble wall created at the 
first-order EWPT, we must know the global structure of the effective 
potential at $T_C$. As an example, we show the contour plot of the 
effective potential at $T=0$ and $T=T_C$ for the case of $\tan\b=5$, 
$m_3^2=4624\mbox{GeV}$ and $m_{\st_R}=0$ in 
Fig.~\ref{fig:Veff-contour}.
This shows that two degenerate minima at $T_C$ is connected by a
valley with almost constant $\tan\b(T_C)\simeq6$.
This is also the case for the other sets of the parameters we studied.
Some of the baryogenesis scenarios based on the MSSM requires that
$\tan\b$ varies spatially around the bubble wall. But this result
implies that $\tan\b$ remains to be almost constant around the wall.
\subsection{$\mib{CP}$-violating case}
In the MSSM, there are many sources of $CP$ violation other than
the phase in the CKM matrix. Among them are relative phases of the 
complex parameters $\mu$, $A_t$, $M_2$ and $M_1$. These also induce
$CP$ violation in the Higgs sector, which is the relative phase $\theta$
of the VEVs of the two Higgs doublets.
This $\theta$ together with the other $CP$-violating phases affect
such observables as the electric dipole moment of the neutron.
Hence knowledge of $\theta$ is necessary to find bounds on the phases
of the complex parameters in the MSSM.
$\theta={\rm Arg}(v_2+iv_3)$ in the gauge with $v_4=0$ is determined 
by minimizing the effective potential at $T=0$.
Even when all the parameters are real, the effective potential could
have $CP$-violating vacuum. This is know as the spontaneous $CP$ 
violation, but in the MSSM it inevitably accompanies a too light 
scalar\cite{Maekawa}. We found that this is the case. Indeed
if we take $\tan\b=5 $ and $m_{\st_R}=0$ and tune 
$m_3^2=2113.3\mbox{GeV}^2$, the effective potential has two 
degenerate minima which correspond to $\theta=\pm0.318$ and 
$\tan\b=4.917$.
Then the lightest scalar mass is $12.9\mbox{GeV}$, which differs
from any mass from the mass formulas because of large mixing of the 
$CP$ eigenstates.\par
\begin{figure}
 \epsfysize=6cm
 \centerline{\epsfbox{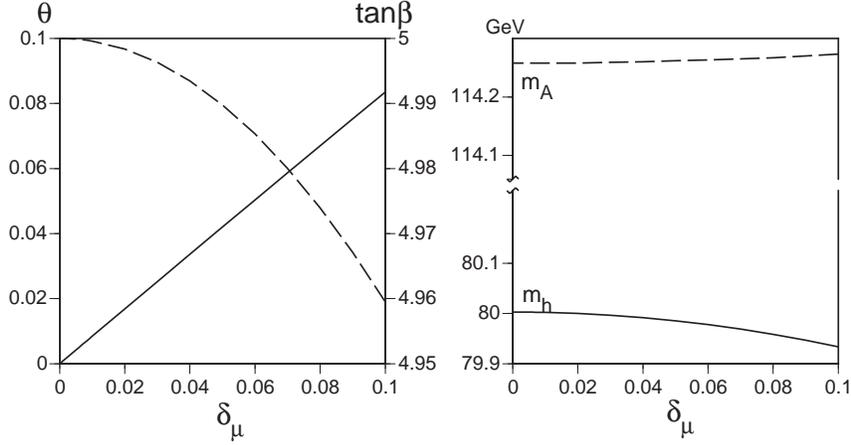}}
 \caption{Dependence of $\theta$ (solid curve), $\tan\b$ (dashed 
 curve) and Higgs masses on $\delta_\mu$.}
 \label{fig:del-mu-dependence}
\end{figure}

Now we study the effect of explicit $CP$ violation in the complex 
parameters on the $CP$ violation in the Higgs sector.
As the first example, we take $\tan\b=5$, $m_{\st_R}=0$, 
$m_3^2=4624\mbox{GeV}^2$, $\mu=-300\cdot e^{i\delta_\mu}$ and the remaining 
parameters are set to be the values adopted in the previous subsection.
For nonzero $\delta_\mu$, $\theta$ have nonzero value and the scalar 
and pseudoscalar mixes to form the mass eigenstates.
By the numerical method, the minimum of the effective potential were 
searched and the second derivatives at the minimum was evaluated to 
calculate the masses of the Higgs bosons, for $0\le\delta_\mu\le0.1$.
Dependence of $\theta$, $\tan\b$ and masses of two light bosons
on $\delta_\mu$ is depicted in Fig.~\ref{fig:del-mu-dependence}.
Within this range of $\delta_\mu$, the derivation of the masses
from the values at $\delta_\mu$ is negligible.
The induced $\theta$ is the same order and has the same sign as
$\delta_\mu$. By linearly fitting, we find
$\theta = 0.8265\cdot \delta_\mu$.
By redefining the fields, we find that the physical $CP$-violating
phase in the mass matrices (\ref{eq:stop-mass-matrix}), 
(\ref{eq:chargino-mass-matrix}) and (\ref{eq:neutralino-mass-matrix}) 
is $\delta_\mu+\theta$.
Hence $\delta_\mu$ enhances the magnitude of $CP$ violation.\par
\begin{figure}
 \epsfysize=6cm
 \centerline{\epsfbox{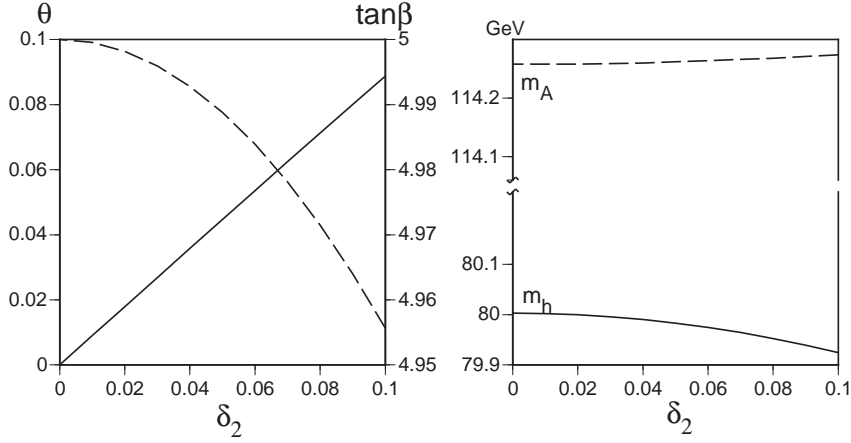}}
 \caption{Dependence of $\theta$ (solid curve), $\tan\b$ (dashed 
 curve) and Higgs masses on $\delta_2$.}
 \label{fig:del2-dependence}
\end{figure}

As a second example, we put $M_2=M_1=350\cdot e^{i\delta_2}\mbox{GeV}$ and
all the rest parameters are taken to be real and set to the same values
as the previous example. 
Dependence of $\theta$, $\tan\b$ and masses of two light bosons
on $\delta_\mu$ is shown in Fig.~\ref{fig:del2-dependence}.
The induced $CP$ phase is fitted to $\theta = 0.8885\cdot \delta_2$, which 
has the same sign as the original $\delta_2$.
Since the physical $CP$ phase in the mass matrix 
(\ref{eq:chargino-mass-matrix}) is $\delta_2+\theta$, the $CP$-violating
phase is enhanced by the radiative corrections.\par
\begin{figure}
 \epsfysize=6cm
 \centerline{\epsfbox{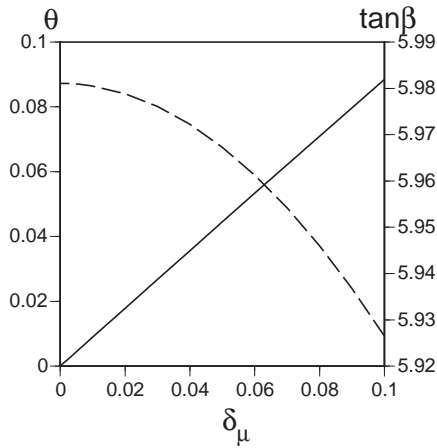}}
 \caption{Dependence of $\theta$ (solid curve), $\tan\b$ (dashed 
 curve) at $T=92\mbox{GeV}$.}
 \label{fig:del-mu-dependence-T}
\end{figure}

In the scenario of the electroweak baryogenesis, the $CP$ violation
around the bubble wall is a key ingredient and it is determined by
solving the equations of motion with the effective potential at $T_C$.
Then the VEVs in the two degenerate minima provide the boundary 
conditions to these equations. 
Although spontaneous $CP$ violation at $T\simeq T_C$ could occur,
it would accompany a too light scalar at zero temperature.
Any way, an explicit $CP$ violation is necessary to resolve degeneracy
in energy of the $CP$-conjugate pair of the bubble walls, otherwise
no net baryon asymmetry would survive the EWPT.
For the same parameters as the zero temperature case, we plot in
Fig.~\ref{fig:del-mu-dependence-T} 
dependences of the induced $CP$ violating phase $\theta$ and $\tan\b$ on
the phase of the $\mu$-parameter. 
The behaviors of $\theta$ and $\tan\b$ are almost the same as those
at zero temperature, but $\theta = 0.8862\cdot\delta_\mu$.
\section{Discussion}
We have studied the masses of the neutral Higgs bosons and the 
electroweak phase transition of the MSSM by use of the one-loop
effective potential. For the parameters we adopted, the contributions 
from the charginos and neutralinos are shown not to be negligible.
We have found that the EWPT is of so strongly first order that
the sphaleron process after it decouples, for 
$m_{\st_R}\ltsim75\mbox{GeV}$ when $\tan\b=2$, $m_h=62.8\mbox{GeV}$ 
and $m_A=239\mbox{GeV}$, for 
$m_{\st_R}\ltsim50\mbox{GeV}$ when $\tan\b=5$, $m_h=62.8\mbox{GeV}$ 
and $m_A=70\mbox{GeV}$, for 
$m_{\st_R}\ltsim53\mbox{GeV}$ when $\tan\b=5$, $m_h=80\mbox{GeV}$ 
and $m_A=114\mbox{GeV}$, for 
$m_{\st_R}\ltsim46.7\mbox{GeV}$ when $\tan\b=20$, $m_h=62.7\mbox{GeV}$ 
and $m_A=62.6\mbox{GeV}$, and for 
$m_{\st_R}\ltsim46.8\mbox{GeV}$ when $\tan\b=20$, $m_h=80\mbox{GeV}$ 
and $m_A=81\mbox{GeV}$. These bounds on $m_{\st_R}$ almost correspond
to a bound on the lighter stop mass $m_{\st_1}\le m_t$.
For the parameters which permit strongly first-order EWPT, we have 
investigated $\tan\b$ at the transition temperature. It receives
larger temperature-corrections for larger $\tan\b$ at zero 
temperature, which corresponds to smaller Yukawa coupling of the top 
quark. This suggests importance of finite-temperature contributions
from the particles other than the top quark and squarks.
We also studied $CP$ violation in the Higgs sector, which is 
characterized by the relative phase $\theta$ of the expectation values
of the two Higgs doublets. 
As is well known, the spontaneous mechanism to generate $\theta$ 
accompanies a too light scalar. An explicit $CP$ violation in the 
complex parameters induces $\theta$ of the same order and sign as 
itself, through radiative and finite-temperature corrections.
This implies that the physical phases in the mass matrices of the 
chargino, neutralino and stop are enhanced by the complex phases 
which are originally contained in these matrices.
Hence one must take this effect into account to find bounds on the 
parameters in the MSSM obtained from such data as the neutron EDM.\par
Some of the mechanism of electroweak baryogenesis in the MSSM requires
$\tan\b$ to vary spatially. But at the transition temperature, it stays
almost constant at $\tan\b(T_C)$. Then viable scenarios of the 
electroweak baryogenesis rely on spatially varying $\theta$ and/or
$\absv{\mib v}$ in the presence explicit $CP$ violation.
The spatial dependence of $\theta$ and $\absv{\mib v}$ around the 
bubble wall created at the EWPT is examined in \cite{explicitCPV}.
The values of these variables in the broken phase at $T_C$ obtained 
here will serve as the boundary conditions to the dynamical 
equations for $(\theta(x),\absv{{\mib v}(x)})$.
These functions in the MSSM will be studied elsewhere\cite{FKOTprep}.\par
In this paper, we extensively used the one-loop effective potential
for self-containedness.
Extension to the two-loop resummed potential would be straightforward.
The two-loop resummed potential without the contributions from the
charginos and neutralinos yields strongly first-order EWPT for
a wider range of parameters than the corresponding one-loop 
potential\cite{ewpt-2loop-mssm}. We expect that if the higher-order 
effects are taken into account, the effective potential including
the contributions from all the particles in the MSSM will provide 
strongly first-order EWPT for a broader region in the parameter space
than that investigated here.
\section*{Acknowledgments}
The author thanks Y.~Okada for let him know recent papers on the
Higgs mass in the MSSM. He also thanks A.~Kakuto, S.~Otsuki and 
F.~Toyoda for discussions on electroweak baryogenesis.
This work is supported in part by Grant-in-Aid for Scientific
Research on Priority Areas (Physics of $CP$ violation, No.10140220)
and No.09740207 from the Ministry of Education, Science, 
and Culture of Japan.
\appendix
\section{Derivatives of the Effective Potential}
We present the formulas for the first and second derivatives of the 
effective potential evaluated at a $CP$-conserving vacuum.
Since all the parameters are assumed to be real, $CP$ is conserved so 
that $v_3=v_4=0$ at the vacuum. But we retain $v_3$ to derive the mass
of the $CP$-odd scalar according to (\ref{eq:def-mA}).
To see contribution from each species, we give the derivatives
of the correction to the effective potential from each particle.
\subsection{gauge bosons}
The contributions to the effective potential is given by 
(\ref{eq:deltaV-gauge}). Its first derivatives at the vacuum are
\begin{eqnarray}
 {1\over{v_1}}{{\del\Delta_g V({\mib v})}\over{\del v_1}} &=&
 {1\over{v_2}}{{\del\Delta_g V({\mib v})}\over{\del v_2}} \nonumber\\
 &=&
 6\cdot{{g_2^2}\over2}{{m_W^2({\mib v})}\over{32\pi^2}}
 \left(\log{{m_W^2({\mib v})}\over{M_{\rm ren}^2}}-1\right) +
 3\cdot{{g_2^2+g_1^2}\over2}{{m_Z^2({\mib v})}\over{32\pi^2}}
 \left(\log{{m_Z^2({\mib v})}\over{M_{\rm ren}^2}}-1\right).\nonumber\\
                            \label{eq:d1-V-gauge}
\end{eqnarray}
It is straightforward to calculate the second derivatives:
\begin{eqnarray}
 v_1{\del\over{\del v_1}}
 \left({1\over{v_1}}{{\del\Delta_g V({\mib v})}\over{\del v_1}}\right) &=&
 {{3\cos^2\b}\over{8\pi^2v_0^2}}\left[
  2 m_W^4\log{{m_W^2({\mib v})}\over{M_{\rm ren}^2}} +
  m_Z^4\log{{m_Z^2({\mib v})}\over{M_{\rm ren}^2}}
                               \right],  \label{eq:d11-V-gauge}\\
 v_2{\del\over{\del v_2}}
 \left({1\over{v_2}}{{\del\Delta_g V({\mib v})}\over{\del v_2}}\right) &=&
 {{3\sin^2\b}\over{8\pi^2v_0^2}}\left[
  2 m_W^4\log{{m_W^2({\mib v})}\over{M_{\rm ren}^2}} +
  m_Z^4\log{{m_Z^2({\mib v})}\over{M_{\rm ren}^2}}
                               \right],  \label{eq:d22-V-gauge}\\
 {{\del^2\Delta_g V({\mib v})}\over{\del v_1\del v_2}} &=&
 {{3\sin\b\cos\b}\over{8\pi^2v_0^2}}\left[
  2 m_W^4\log{{m_W^2({\mib v})}\over{M_{\rm ren}^2}} +
  m_Z^4\log{{m_Z^2({\mib v})}\over{M_{\rm ren}^2}}
                               \right],  \nonumber\\
 & &                              \label{eq:d12-V-gauge}\\
 {{\del^2\Delta_g V({\mib v})}\over{\del v_3^2}} &=&
 {1\over{v_3}}{{\del\Delta_g V({\mib v})}\over{\del v_3}} =
 {1\over{v_2}}{{\del\Delta_g V({\mib v})}\over{\del v_2}}.
                                         \label{eq:d33-V-gauge}
\end{eqnarray}
\subsection{top quark}
Since the contribution from the top quark (\ref{eq:deltaV-top}) is
independent of $v_1$, any derivative with respect to $v_1$ vanishes.
\begin{eqnarray}
 {1\over{v_1}}{{\del\Delta_g V({\mib v})}\over{\del v_1}} &=& 
 0,     \nonumber\\
 {1\over{v_2}}{{\del\Delta_g V({\mib v})}\over{\del v_2}} &=&
 {1\over{v_3}}{{\del\Delta_g V({\mib v})}\over{\del v_3}} =
 -12\cdot y_t^2{{m_t^2({\mib v})}\over{32\pi^2}}
   \left(\log{{m_t^2({\mib v})}\over{M_{\rm ren}^2}}-1\right),
                            \label{eq:d2-V-top}
\end{eqnarray}
and
\begin{eqnarray}
 v_1{\del\over{\del v_1}}
 \left({1\over{v_1}}{{\del\Delta_t V({\mib v})}\over{\del v_1}}\right) &=&
 {{\del^2\Delta_t V({\mib v})}\over{\del v_1\del v_2}} = 0, \\
 v_2{\del\over{\del v_2}}
 \left({1\over{v_2}}{{\del\Delta_t V({\mib v})}\over{\del v_2}}\right) &=&
 -12\cdot{{4m_t^4({\mib v})}\over{v_0^2\sin^2\b}}
    {1\over{32\pi^2}}\log{{m_t^2({\mib v})}\over{M_{\rm ren}^2}},
                                      \label{eq:d22-V-top}\\
 {{\del^2\Delta_t V({\mib v})}\over{\del v_3^2}} &=&
 {1\over{v_2}}{{\del\Delta_t V({\mib v})}\over{\del v_2}}.
                                         \label{eq:d33-V-top}
\end{eqnarray}
\subsection{top squarks}
The stop contribution is given by (\ref{eq:deltaV-stop}), in which
the mass eigenvalues are
\begin{eqnarray}
 m_{\st_a}^2({\mib v}) &=&
 {{m_{\st_L}^2+m_{\st_R}^2}\over2}+{{g_2^2+g_1^2}\over8}(v_1^2-v_2^2-v_3^2)
 + m_t^2({\mib v}) \nonumber\\
 & &\pm
 \sqrt{
  \left({{m_{\st_L}^2-m_{\st_R}^2}\over2} +
        {{3g_2^2-5g_1^2}\over{24}}(v_1^2-v_2^2-v_3^2)\right)^2 +
  {{y_t^2}\over2}\left[(\mu v_1+A_t v_2)^2+A_t^2v_3^2\right] }.\nonumber\\
 & &
\end{eqnarray}
The first derivatives are given by
\begin{equation}
 {1\over{v_i}}{{\del\Delta_\st V({\mib v})}\over{\del v_i}} =
 6\sum_{a=1,2}{1\over{v_i}}{{\del m_{\st_a}^2({\mib v})}\over{\del v_i}}
  {{m_{\st_a}^2({\mib v})}\over{32\pi^2}}\left(
   \log{{m_{\st_a}^2({\mib v})}\over{M_{\rm ren}^2}} -1 \right),
           \label{eq:di-deltaV-stop}
\end{equation}
where the first derivatives of the stop mass-squared evaluated at 
the vacuum are
\begin{eqnarray}
 {1\over{v_1}}{{\del m_{\st_a}^2}\over{\del v_1}} &=&
 {{2m_t^2}\over{v_0^2\sin^2\b}}\left( {{m_Z^2\sin^2\b}\over{2m_t^2}}\pm 
                      {{n_\st^{(1)}}\over{R_\st}} \right),  
             \label{eq:d1-m-stop}\\
 {1\over{v_2}}{{\del m_{\st_a}^2}\over{\del v_2}} &=&
 {{2m_t^2}\over{v_0^2\sin^2\b}}\left( 1-{{m_Z^2\sin^2\b}\over{2m_t^2}}\pm 
       {{n_\st^{(2)}}\over{R_\st}} \right),    \label{eq:d2-m-stop}\\
 {1\over{v_3}}{{\del m_{\st_a}^2}\over{\del v_3}} &=&
 {{2m_t^2}\over{v_0^2\sin^2\b}}\left( 1-{{m_Z^2\sin^2\b}\over{2m_t^2}}\pm 
 { {-{{8m_W^2-5m_Z^2}\over{6m_t^2}}d_\st\sin^2\b +\half A_t^2} \over
   {R_\st} } \right),                  \label{eq:d3-m-stop}
\end{eqnarray}
with
\begin{eqnarray}
 d_\st &=& 
 {{m_{\st_L}^2-m_{\st_R}^2}\over2}+{{8m_W^2-5m_Z^2}\over6}\cos(2\b),
                    \nonumber\\
 n_\st^{(1)} &=&
 {{8m_W^2-5m_Z^2}\over{6m_t^2}}d_\st\sin^2\b+\half\mu(\mu+A_t\tan\b),
                    \nonumber\\
 n_\st^{(2)} &=&
 -{{8m_W^2-5m_Z^2}\over{6m_t^2}}d_\st\sin^2\b+\half A_t(\mu\cot\b+A_t),
                    \nonumber\\
 R_\st &=& \sqrt{d_\st^2 + m_t^2(\mu\cot\b+A_t)^2}.
                    \label{eq:def-d-R-stop}
\end{eqnarray}
The general expression for the second derivatives is given by
\begin{eqnarray}
 v_j{\del\over{\del v_i}}\left({1\over{v_j}}
  {{\del\Delta_\st V({\mib v})}\over{\del v_j}}\right) &=&
 6\sum_{a=1,2}\Biggl[
  v_j{\del\over{\del v_i}}\left({1\over{v_j}}
    {{\del m_{\st_a}^2({\mib v})}\over{\del v_j}}\right)
  {{m_{\st_a}^2({\mib v})}\over{32\pi^2}}\left(
   \log{{m_{\st_a}^2({\mib v})}\over{M_{\rm ren}^2}} -1 \right) \nonumber\\
 & &\qquad\qquad\qquad +
  {{\del m_{\st_a}^2({\mib v})}\over{\del v_i}}
  {{\del m_{\st_a}^2({\mib v})}\over{\del v_j}}{1\over{32\pi^2}}
   \log{{m_{\st_a}^2({\mib v})}\over{M_{\rm ren}^2}} \Biggr]
                                                 \label{eq:ddV-by-ddm}
\end{eqnarray}
The relevant second derivatives to this expression are
\begin{eqnarray}
 v_1{\del\over{\del v_1}}\left({1\over{v_1}}
      {{\del m_{\st_a}^2}\over{\del v_1}}\right)
 &=&
 \pm{{\cos^2\b}\over{v_0^2}}\left[
  {{ \left({{8m_W^2-5m_Z^2}\over3}\right)^2 -
     {{\mu A_t m_t^2}\over{\sin\b\cos^3\b}} }\over{R_\st} } -
 {{\left(n_\st^{(1)} \right)^2}\over{R_\st^3} } \right],
                            \label{eq:d11-m-stop}\\
 v_2{\del\over{\del v_2}}\left({1\over{v_2}}
      {{\del m_{\st_a}^2}\over{\del v_2}}\right)
 &=&
 \pm{{\sin^2\b}\over{v_0^2}}\left[
  {{ \left({{8m_W^2-5m_Z^2}\over3}\right)^2 -
     {{\mu A_t m_t^2\cos\b}\over{\sin^5\b}} }\over{R_\st} } -
 {{\left(n_\st^{(2)}\right)^2}\over{R_\st^3} } \right],
                             \label{eq:d22-m-stop}\\
 {{\del^2 m_{\st_a}^2}\over{\del v_1\del v_2}}
 &=&
 \pm{{\sin\b\cos\b}\over{v_0^2}}\left[
  {{ -\left({{8m_W^2-5m_Z^2}\over3}\right)^2 +
     {{\mu A_t m_t^2}\over{\sin^3\b\cos\b}} }\over{R_\st} } -
  {{n_\st^{(1)}n_\st^{(2)}}\over{R_\st^3}} \right],
                             \label{eq:d12-m-stop}\\
 v_3{\del\over{\del v_3}}\left({1\over{v_3}}
      {{\del m_{\st_a}^3}\over{\del v_3}}\right) &=& 0.
\end{eqnarray}
\subsection{charginos}
The derivatives of (\ref{eq:deltaV-chargino}) are evaluated in the 
same manner as the case of the stop. The mass-squared eigenvalues are
\begin{eqnarray}
 m_{\chi^\pm_a}^2({\mib v}) &=&
 m_W^2({\mib v}) +{{M_2^2+\mu^2}\over2}  \nonumber\\
 & &\pm
 \sqrt{\left({{M_2^2-\mu^2}\over2}+{{g_2^2}\over4}(v_1^2-v_2^2-v_3^2)
       \right)^2 + 
      {{g_2^2}\over2}\left[(\mu v_1+M_2v_2)^2+M_2^2v_3^2\right]}.
         \nonumber\\   \label{eq:chargino-mass}
\end{eqnarray}
The first derivatives of the corrections to the effective potential
is given by a similar expression to (\ref{eq:di-deltaV-stop}), in 
which the relevant derivatives of the masses-squared are
\begin{eqnarray}
 {1\over{v_1}}{{\del m_{\chi^\pm_a}^2}\over{\del v_1}} &=&
 {{2m_W^2}\over{v_0^2}}\left(1\pm {{n_\chi^{(1)}}\over{R_\chi}}\right),
                           \label{eq:d1-m-chargino}\\
 {1\over{v_2}}{{\del m_{\chi^\pm_a}^2}\over{\del v_2}} &=&
 {{2m_W^2}\over{v_0^2}}\left(1\pm {{n_\chi^{(2)}}\over{R_\chi}}\right),
                          \label{eq:d2-m-chargino}\\
 {1\over{v_3}}{{\del m_{\chi^\pm_a}^2}\over{\del v_3}} &=&
 {{2m_W^2}\over{v_0^2}}\left( 1 \pm
  { {{{M_2^2+\mu^2}\over2}-m_W^2\cos(2\b)} \over{R_\chi}}\right),
                          \label{eq:d3-m-chargino}
\end{eqnarray}
where
\begin{eqnarray}
 n_\chi^{(1)} &=&
 {{M_2^2+\mu^2}\over2}+m_W^2\cos(2\b)+\mu M_2\tan\b,  \nonumber\\
 n_\chi^{(2)} &=&
 {{M_2^2+\mu^2}\over2}-m_W^2\cos(2\b)+\mu M_2\cot\b,  \nonumber\\
 R_\chi &=&
 \sqrt{\left({{M_2^2-\mu^2}\over2}+m_W^2\cos(2\b)\right)^2 +
       2m_W^2(M_2\sin\b+\mu\cos\b)^2}. \label{eq:def-n-R-chargino}
\end{eqnarray}
The second derivatives have the same form as (\ref{eq:ddV-by-ddm})
except for the overall coefficient. The relevant second derivatives 
are given by
\begin{eqnarray}
 v_1{\del\over{\del v_1}}
  \left({1\over{v_1}}{{\del m_{\chi^\pm_a}^2}\over{\del v_1}}\right)
 &=& \pm
 {{2m_W^2}\over{v_0^2}}\left[
  {{2m_W^2\cos^2\b-\mu M_2\tan\b}\over{R_\chi}} -
  {{2m_W^2\left(n_\chi^{(1)}\right)^2\cos^2\b}\over{R_\chi^3}}
                       \right],   \nonumber\\
 & &                  \label{eq:d11-m-chargino} \\
 v_2{\del\over{\del v_2}}
  \left({1\over{v_2}}{{\del m_{\chi^\pm_a}^2}\over{\del v_2}}\right)
 &=& \pm
 {{2m_W^2}\over{v_0^2}}\left[
   {{2m_W^2\sin^2\b-\mu M_2\cot\b}\over{R_\chi}} -
   {{2m_W^2\left(n_\chi^{(1)}\right)^2\sin^2\b}\over{R_\chi^3}}
                       \right],    \nonumber\\
 & &                  \label{eq:d22-m-chargino}\\
 {{\del^2 m_{\chi^\pm_a}^2}\over{\del v_1\del v_2}}
 &=& \pm
 {{2m_W^2}\over{v_0^2}}\left[
  {{-2m_W^2\sin\b\cos\b+\mu M_2}\over{R_\chi}} -
  {{2m_W^2 n_\chi^{(1)} n_\chi^{(2)}\sin\b\cos\b}\over{R_\chi^3}}
                       \right],   \nonumber\\
 & &                              \label{eq:d12-m-chargino}\\
 v_3{\del\over{\del v_3}}
  \left({1\over{v_3}}{{\del m_{\chi^\pm_a}^2}\over{\del v_3}}\right)
 &=& 0. 
\end{eqnarray}
\subsection{neutralinos}
Although the neutralino contribution to the effective potential is 
given by (\ref{eq:deltaV-neutralino}) in terms of the mass eigenvalues,
it is difficult to work out its derivatives, since the eigenvalues 
have complicated forms. To avoid this complexity, we return to the 
original form of the neutralino contribution:
\begin{equation}
 \Delta_{\chi^0} V({\mib v}) =
 {i\over2}\int_k \Tr\,\log\left[ D_{\chi^0}^{-1}(k;{\mib v})\right],
                      \label{eq:def-chi0-V}
\end{equation}
where $\int_k$ denotes the integral over the Minkowskian momentum,
the trace is taken over the index of 4-dimensional internal 
space and the spinor index, and $D_{\chi^0}$ is the four-component
Dirac operator defined by
\begin{equation}
 D_{\chi^0}^{-1}(k;{\mib v}) =
 \kslash - M_{\chi^0}{{1-\g_5}\over2} 
         - M_{\chi^0}^\dagger{{1+\g_5}\over2}.
\end{equation}
Here the mass matrix $M_{\chi^0}$ is defined by 
(\ref{eq:neutralino-mass-matrix}). The first and second derivatives of
(\ref{eq:def-chi0-V}) have the forms of
\begin{eqnarray}
 {{\del\Delta_{\chi^0}V({\mib v})}\over{\del v_i}} &=&
 {i\over2}\int_k \Tr\left[ D_{\chi^0}(k;{\mib v})
   {{\del D_{\chi^0}^{-1}(k;{\mib v})}\over{\del v_i}}\right],\nonumber\\
 {{\del^2\Delta_{\chi^0}V({\mib v})}\over{\del v_i\del v_j}} &=&
 -{i\over2}\int_k \Tr\left[ D_{\chi^0}(k;{\mib v})
   {{\del D_{\chi^0}^{-1}(k;{\mib v})}\over{\del v_i}} D_{\chi^0}(k;{\mib v})
   {{\del D_{\chi^0}^{-1}(k;{\mib v})}\over{\del v_j}}\right]. 
\end{eqnarray}
The integrand of the first derivative evaluated at the vacuum have
the following compact form:
\begin{equation}
 {1\over{v_i}}\Tr\left[ D_{\chi^0}(k;{\mib v})
    {{\del D_{\chi^0}^{-1}(k;{\mib v})}\over{\del v_i}}\right] =
 -{{2m_Z^2}\over{v_0^2}}\Tr\left[
   {{\kslash+\mu_i}\over{k^2-\mu^2}}
   {{D_1(k)}\over{1 - m_Z^2D_1(k)D_2(k)}} \right],  \label{eq:di-Tr-DD}
\end{equation}
where
\begin{equation}
 D_1(k) = {{\cos^2\th_W}\over{\kslash-M_2}}+{{\sin^2\th_W}\over{\kslash-M_1}},
 \qquad
 D_2(k) = {{\kslash+\mu\sin(2\b)}\over{k^2-\mu^2}},
\end{equation}
and
\begin{equation}
 \mu_1 \equiv \mu\tan\b,\qquad \mu_2 \equiv \mu\cot\b.
\end{equation}
The integrands of the second derivatives are reduced to
\begin{eqnarray}
 & &
 \Tr\left[ D_{\chi^0}(k;{\mib v})
   {{\del D_{\chi^0}^{-1}(k;{\mib v})}\over{\del v_1}}D_{\chi^0}(k;{\mib v})
   {{\del D_{\chi^0}^{-1}(k;{\mib v})}\over{\del v_1}}\right] \nonumber\\
 &=&
 {{2m_Z^2}\over{v_0^2}} \Tr\left[
  {\kslash\over{k^2-\mu^2}}{{D_1(k)}\over{1 - m_Z^2D_1(k)D_2(k)}}
                           \right]  \nonumber\\ 
 & &\qquad\qquad +
 {{4m_Z^4\cos^2\b}\over{v_0^2}} \Tr\left[
   \left( {{\kslash+\mu\tan\b}\over{k^2-\mu^2}}
          {{D_1(k)}\over{1 - m_Z^2D_1(k)D_2(k)}} \right)^2 \right],
                           \label{eq:d11-DDDD}\\
 & &
 \Tr\left[ D_{\chi^0}(k;{\mib v})
   {{\del D_{\chi^0}^{-1}(k;{\mib v})}\over{\del v_2}}D_{\chi^0}(k;{\mib v})
   {{\del D_{\chi^0}^{-1}(k;{\mib v})}\over{\del v_2}}\right] \nonumber\\
 &=&
 {{2m_Z^2}\over{v_0^2}} \Tr\left[
  {\kslash\over{k^2-\mu^2}}{{D_1(k)}\over{1 - m_Z^2D_1(k)D_2(k)}}
                            \right]  \nonumber\\ 
 & &\qquad\qquad + 
 {{4m_Z^4\sin^2\b}\over{v_0^2}} \Tr\left[
   \left({{\kslash+\mu\cot\b}\over{k^2-\mu^2}}
          {{D_1(k)}\over{1 - m_Z^2D_1(k)D_2(k)}} \right)^2 \right],
                         \label{eq:d22-DDDD}\\
 & &
 \Tr\left[ D_{\chi^0}(k;{\mib v})
   {{\del D_{\chi^0}^{-1}(k;{\mib v})}\over{\del v_1}}D_{\chi^0}(k;{\mib v})
   {{\del D_{\chi^0}^{-1}(k;{\mib v})}\over{\del v_2}}\right] \nonumber\\
 &=&
 {{2m_Z^2}\over{v_0^2}} \Tr\left[
    {\mu\over{k^2-\mu^2}}{{D_1(k)}\over{1 - m_Z^2D_1(k)D_2(k)}}
                           \right]  \nonumber\\ 
 &+& 
 {{4m_Z^4\sin\b\cos\b}\over{v_0^2}} \Tr\left[
   {{\kslash+\mu\tan\b}\over{k^2-\mu^2}}{{\kslash+\mu\cot\b}\over{k^2-\mu^2}}
   \left({{D_1(k)}\over{1 - m_Z^2D_1(k)D_2(k)}} \right)^2 \right],
                        \label{eq:d12-DDDD}\\
 & &
 \Tr\left[ D_{\chi^0}(k;{\mib v})
   {{\del D_{\chi^0}^{-1}(k;{\mib v})}\over{\del v_3}}D_{\chi^0}(k;{\mib v})
   {{\del D_{\chi^0}^{-1}(k;{\mib v})}\over{\del v_3}}\right] \nonumber\\
 &=&
 {{2m_Z^2}\over{v_0^2}}\Tr\left[
    {\kslash\over{k^2-\mu^2}}{{D_1(k)}\over{1 - m_Z^2D_1(k)D_2(k)}}
                          \right]       \nonumber\\
 &-&
 {{4m_Z^4\cos^2\b}\over{v_0^2}}{{\mu^2}\over{(k^2-\mu^2)^2}}
 \Tr\left[{{D_1(k)}\over{1 - m_Z^2D_1(k)D_2(k)}}
          {{D_1(-k)}\over{1 - m_Z^2D_1(-k)D_2(-k)}} \right].
                             \label{eq:d33-DDDD}
\end{eqnarray}
\par
For the special case of $M_2=M_1$ which is extensively investigated 
in this paper, we have the following formulas for the relevant derivatives
to the masses of the neutral Higgs bosons.
\begin{eqnarray}
 & &
 {1\over{v_i}}{{\del\Delta_{\chi^0}V({\mib v})}\over{\del v_i}}\nonumber\\
 &=&
 {{m_Z^2}\over{4\pi^2v_0^2}}\Biggl\{
  -(\mu^2+m_Z^2)\left(\log{{\mu^2+m_Z^2}\over{M_{\rm ren}^2}}-1\right) +
   M_2(M_2+\mu_i)L(M_2^2,\mu^2+m_Z^2)  \nonumber\\
 & &\qquad
 +\left(1+{{\mu\sin(2\b)}\over{M_2}}\right) m_Z^2
  \left[-\left(2+{{\mu_i}\over{M_2}}\right)
        F_1\left({\mu\over{M_2}},{{m_Z}\over{M_2}},\tan\b\right) 
     \right.\nonumber\\
 & &\qquad\qquad\qquad\qquad\qquad\qquad\left. +
        2\left(1+{{\mu_i}\over{M_2}}\right)
        F_2\left({\mu\over{M_2}},{{m_Z}\over{M_2}},\tan\b\right)
  \right]    \nonumber\\
 & &\qquad
 +\left(1+{{\mu\sin(2\b)}\over{M_2}}\right)^2{{m_Z^4}\over{M_2^2}}
  \left[ F_3\left({\mu\over{M_2}},{{m_Z}\over{M_2}},\tan\b\right)
 -\left(1+{{\mu_i}\over{M_2}}\right)
     F_4\left({\mu\over{M_2}},{{m_Z}\over{M_2}},\tan\b\right) \right\},
                       \nonumber\\
 & &                   \label{eq:di-V-chi-0}\\
 & &
 -{1\over{v_1}}{{\del\Delta_{\chi^0}V({\mib v})}\over{\del v_1}}
 +{{\del^2\Delta_{\chi^0}V({\mib v})}\over{\del v_1^2}}  \nonumber\\
 &=&
 {{m_Z^2}\over{4\pi^2v_0^2}}\Biggl\{
  (2m_Z^2\cos^2\b-\mu_1M_2)L(M_2^2,\mu^2+m_Z^2)   \nonumber\\
 & &\qquad
 +\left[{{\mu_1}\over{M_2}}\left(1+{{\mu\sin(2\b)}\over{M_2}}\right)
       -{{\mu^2+m_Z^2\cos^2\b}\over{M_2^2}}\right] m_Z^2
     F_1\left({\mu\over{M_2}},{{m_Z}\over{M_2}},\tan\b\right) \nonumber\\
 & &\qquad
 +2\left(1+{{\mu\sin(2\b)}\over{M_2}}\right)
   \left({{2m_Z^2\cos^2\b}\over{M_2^2}}-{{\mu_1}\over{M_2}}\right) m_Z^2
     F_2\left({\mu\over{M_2}},{{m_Z}\over{M_2}},\tan\b\right) \nonumber\\
 & &\qquad
 +\left(1+{{\mu\sin(2\b)}\over{M_2}}\right)^2
  \left({{\mu_1}\over{M_2}}-{{2m_Z^2\cos^2\b}\over{M_2^2}}\right)
  {{m_Z^4}\over{M_2^2}}
     F_4\left({\mu\over{M_2}},{{m_Z}\over{M_2}},\tan\b\right) \nonumber\\
 & &\qquad
 -4m_Z^2\cos^2\b G_{11}\left({\mu\over{M_2}},{{m_Z}\over{M_2}},\tan\b\right)
   \Biggr\},          \label{eq:d11-V-chi-0}\\
 & &
 -{1\over{v_2}}{{\del\Delta_{\chi^0}V({\mib v})}\over{\del v_2}}
 +{{\del^2\Delta_{\chi^0}V({\mib v})}\over{\del v_2^2}}  \nonumber\\
 &=&
 {{m_Z^2}\over{4\pi^2v_0^2}}\Biggl\{
  (2m_Z^2\sin^2\b-\mu_2M_2)L(M_2^2,\mu^2+m_Z^2)   \nonumber\\
 & &\qquad
 +\left[{{\mu_2}\over{M_2}}\left(1+{{\mu\sin(2\b)}\over{M_2}}\right)
       -{{\mu^2+m_Z^2\sin^2\b}\over{M_2^2}}\right] m_Z^2
     F_1\left({\mu\over{M_2}},{{m_Z}\over{M_2}},\tan\b\right) \nonumber\\
 & &\qquad
 +2\left(1+{{\mu\sin(2\b)}\over{M_2}}\right)
   \left({{2m_Z^2\sin^2\b}\over{M_2^2}}-{{\mu_2}\over{M_2}}\right) m_Z^2
     F_2\left({\mu\over{M_2}},{{m_Z}\over{M_2}},\tan\b\right) \nonumber\\
 & &\qquad
 +\left(1+{{\mu\sin(2\b)}\over{M_2}}\right)^2
  \left({{\mu_2}\over{M_2}}-{{2m_Z^2\sin^2\b}\over{M_2^2}}\right)
  {{m_Z^4}\over{M_2^2}}
     F_4\left({\mu\over{M_2}},{{m_Z}\over{M_2}},\tan\b\right) \nonumber\\
 & &\qquad
 -4m_Z^2\sin^2\b G_{22}\left({\mu\over{M_2}},{{m_Z}\over{M_2}},\tan\b\right)
   \Biggr\},           \label{eq:d22-V-chi-0}\\
 & &
 {{\del^2\Delta_{\chi^0}V({\mib v})}\over{\del v_1\del v_2}}  \nonumber\\
 &=&
 {{m_Z^2}\over{4\pi^2v_0^2}}\Biggl\{
  (\mu M_2+m_Z^2\sin(2\b))L(M_2^2,\mu^2+m_Z^2)   \nonumber\\
 & &\qquad
 -\left[{\mu\over{M_2}}\left(1+{{\mu\sin(2\b)}\over{M_2}}\right)
       +{{(2\mu^2+m_Z^2)\sin(2\b)}\over{M_2^2}}\right] m_Z^2
     F_1\left({\mu\over{M_2}},{{m_Z}\over{M_2}},\tan\b\right) \nonumber\\
 & &\qquad
 +2\left(1+{{\mu\sin(2\b)}\over{M_2}}\right)
   \left({\mu\over{M_2}}+{{m_Z^2\sin(2\b)}\over{M_2^2}}\right) m_Z^2
     F_2\left({\mu\over{M_2}},{{m_Z}\over{M_2}},\tan\b\right) \nonumber\\
 & &\qquad
 -\left(1+{{\mu\sin(2\b)}\over{M_2}}\right)^2
  \left({{\mu}\over{M_2}}+{{m_Z^2\sin(2\b)}\over{M_2^2}}\right)
  {{m_Z^4}\over{M_2^2}}
     F_4\left({\mu\over{M_2}},{{m_Z}\over{M_2}},\tan\b\right) \nonumber\\
 & &\qquad
 -2m_Z^2\sin(2\b) G_{12}\left({\mu\over{M_2}},{{m_Z}\over{M_2}},\tan\b\right)
   \Biggr\},           \label{eq:d12-V-chi-0}\\
 & &
 -{1\over{v_2}}{{\del\Delta_{\chi^0}V({\mib v})}\over{\del v_2}}
 +{{\del^2\Delta_{\chi^0}V({\mib v})}\over{\del v_3^2}}  \nonumber\\
 &=&
 {{m_Z^2}\over{4\pi^2v_0^2}}\Biggl\{
  -\mu_2 M_2 L(M_2^2,\mu^2+m_Z^2)  
 +{{\mu_2}\over{M_2}} m_Z^2
     F_1\left({\mu\over{M_2}},{{m_Z}\over{M_2}},\tan\b\right) \nonumber\\
 & &\qquad
 -2{{\mu_2}\over{M_2}}\left(1+{{\mu\sin(2\b)}\over{M_2}}\right) m_Z^2
     F_2\left({\mu\over{M_2}},{{m_Z}\over{M_2}},\tan\b\right) \nonumber\\
 & &\qquad
 +{{\mu_2}\over{M_2}}\left(1+{{\mu\sin(2\b)}\over{M_2}}\right)^2
  {{m_Z^4}\over{M_2^2}}
     F_4\left({\mu\over{M_2}},{{m_Z}\over{M_2}},\tan\b\right)
   \Biggr\}.          \label{eq:d33-V-chi-0}
\end{eqnarray}
Here various functions arising from the momentum integrals are 
defined by
\begin{eqnarray}
 L(m_1^2,m_2^2) &=& \left\{
  \begin{array}{ll}
   -\log{{m_1^2}\over{M_{\rm ren}^2}}, &\mbox{for $m_1^2=m_2^2$}\\
   -{{m_1^2}\over{m_1^2-m_2^2}}\log{{m_1^2}\over{M_{\rm ren}^2}}
   +{{m_2^2}\over{m_1^2-m_2^2}}\log{{m_2^2}\over{M_{\rm ren}^2}}+1.
                                    &\mbox{for $m_1^2\not=m_2^2$}
  \end{array}     \right.      \label{eq:def-L}\\
 F_1(a,b,\tan\b) &=&
 \int_0^\infty dx
 {x\over{x(x+a^2+b^2)^2+\left(x+a^2-ab^2\sin(2\b)\right)^2}},
                               \label{eq:def-F1}\\
 F_2(a,b,\tan\b) &=&
 \int_0^\infty dx
  {x\over
   {(x+1)\left[x(x+a^2+b^2)^2+\left(x+a^2-ab^2\sin(2\b)\right)^2\right]}},
                               \label{eq:def-F2}\\
 F_3(a,b,\tan\b) &=&
 \int_0^\infty dx
  {x\over
   {(x+a^2+b^2)
    \left[x(x+a^2+b^2)^2+\left(x+a^2-ab^2\sin(2\b)\right)^2\right]}},
                               \label{eq:def-F3}\\
 F_4(a,b,\tan\b) &=&
 \int_0^\infty dx
  {x\over
   {(x+1)(x+a^2+b^2)
    \left[x(x+a^2+b^2)^2+\left(x+a^2-ab^2\sin(2\b)\right)^2\right]}},
                               \nonumber\\
 & &                           \label{eq:def-F4}\\
 G_{ij}(a,b,\tan\b) &=&
 \int_0^\infty dx
  {{(x-a_ia_j)\left(x+a^2-ab^2\sin(2\b)\right) +
    (a_i+a_j)x(x+a^2+b^2)}\over
   {\left[x(x+a^2+b^2)^2+\left(x+a^2-ab^2\sin(2\b)\right)^2\right]^2}
  },                           \nonumber\\
 & &                           \label{eq:def-Gij}
\end{eqnarray}
where $a_i=a\cdot\mu_i/\mu$.
%
%
%


\baselineskip=13pt
\begin{thebibliography}{99}
\bibitem{reviewEB} For a review see, A.~Cohen, D.~Kaplan and A.~Nelson,
Ann. Rev. Nucl. Part. Sci. {\bf 43} (1993) 27.  \\
K.~Funakubo,\PrTP{96}{96}{475}.
\bibitem{BKS} A.~I.~Bochkarev, S.~V.~Kuzmin and M.~E.~Shaposhnikov,
\PLB{244}{90}275.
\bibitem{EWPT-MSM} For nonperturbative studies on the lattice, see
F.~Csikor, Z.~Fodor and J.~Heitger, hep-lat/9807021 and references therein.
\bibitem{light-stop1} A.~Brignole, J.~R.~Espinosa, M.~Quir\'os and 
F.~Zwirner, \PLB{324}{94}{181}.\\
M.~Carena, M.~Quir\'os and C.~E.~M.~Wagner, \PLB{380}{96}{81}.\\
D.~Delepine, J.-M.~G\'erard, R.~Gonzalez~Filipe, J.~Weyers, 
\PLB{386}{96}{183}.\\
J.~M.~Cline and G.~D.~Moore, hep-ph/9806354. 
\bibitem{higgsmass-mssm1} Y.~Okada, M.~Yamaguchi and T.~Yanagida, 
\PrTP{85}{91}{1}; \PLB{262}{91}{54}.\\
J.~Ellis, G.~Ridolfi and F.~Zwirner, \PLB{257}{91}{83}; 
\PLB{262}{91}{477}.\\
H.~E.~Haber and R.~Hempfling, \PRLet{66}{91}{1815}.\\
J.~R.~Espinosa and M.~Quir\'os, \PLB{266}{91}{389}.\\
D.~M.~Pierce, A.~Papadopoulos and S.~B.~Johnson, \PRLet{68}{92}{3678}.
\bibitem{EWBsusy} A.~G.~Cohen and A.~E~.Nelson, \PLB{297}{92}{111}.\\
P.~Huet and A.~E.~Nelson, \PRD{53}{96}{4578}.\\
M.~Carena, M.~Quir\'os, A.~Riotto, I.~Vilja and C.~E.~M.~Wagner,
\NPB{503}{97}{387}.\\
M.~P.~Worah, \PRLet{79}{97}{3810}.\\
M.~Aoki, N~.Oshimo and A.~Sugamoto, \PrTP{98}{97}{1179}.
\bibitem{higgs-phase} A.~Nelson, D.~Kaplan and A.~Cohen, 
\NPB{373}{92}{453}.\\
K.~Funakubo, A.~Kakuto, S.~Otsuki, K.~Takenaga and F.~Toyoda,
\PRD{50}{94}{1105}.\\
K.~Funakubo, A.~Kakuto, S.~Otsuki and F.~Toyoda, \PrTP{95}{96}{929}.
\bibitem{spontCPV} D.~Comelli, M.~Pietroni and A.~Riotto, 
\NPB{412}{94}{441}.
\bibitem{FKOTT} K.~Funakubo, A.~Kakuto, S.~Otsuki, K.~Takenaga and F.~Toyoda,
\PrTP{95}{96}{771}.
\bibitem{FKOTmssm} K.~Funakubo, A.~Kakuto, S.~Otsuki and F.~Toyoda,
\PrTP{99}{98}{1045}.
\bibitem{ewpt-2loop-mssm} J.~R.~Espinosa, \NPB{475}{96}{273}.\\
B.~de~Carlos and J.~R.~Espinosa, \NPB{503}{97}{24}.
\bibitem{higgsmass-mssm2} A.~Dabelstein, Z. Phys. {\bf C67} (1995) 495.
\bibitem{higgsmass-mssm3} S.~Heinemeyer, W.~Hollik and G.~Weiglein,
hep-ph/9807423.\\
R.-J.~Zhang, hep-ph/9808299.
\bibitem{DJ} L.~Dolan and R.~Jackiw, \PRD{9}{74}{3320}.
\bibitem{explicitCPV} K.~Funakubo, A.~Kakuto, S.~Otsuki and F.~Toyoda, 
\PrTP{96}{96}{771}; \PrTP{98}{97}{427}.
\bibitem{num-recipe} W.~H.~Press, B.~P.~Flannery, S.~A.~Teukovsky and
W.~T.~Vetterling, {\sl Numerical Recipes in C} (Cambrigde University 
Press, 1988).
\bibitem{sph-decouple} M.~E.~Shaposhnikov, JETP Lett. {\bf 44} (1986) 
465; \NPB{287}{87}{757}.
\bibitem{pd} Particle Data Group, Eur. Phys. J. {\bf C3} (1998) 1.
\bibitem{Barger} V.~Barger, hep-ph/9808354.
\bibitem{Maekawa} N.~Maekawa, \PLB{282}{92}{387}; Nucl. Phys. Suppl.
{\bf 37A} (1994) 191.
\bibitem{FKOTprep} K.~Funakubo, A.~Kakuto, S.~Otsuki and F.~Toyoda,
in preparation.
%
\end{thebibliography}
\end{document}